\documentclass[twocolumn]{aastex631}

\usepackage[utf8]{inputenc}
\usepackage{textgreek}
\usepackage[english]{babel}
\usepackage{tabularx} 
\usepackage{booktabs}
\usepackage{natbib}
\usepackage{amsmath,amssymb,amsfonts}%
\usepackage{amsthm}%
\usepackage{mathrsfs}%
\usepackage{graphicx}
\usepackage{cuted}
\usepackage{url}
\usepackage{hyperref} 

\usepackage{booktabs}
\usepackage{longtable}
\usepackage{array}

\begin{document}

\title{Model-Driven Stellar Spectroscopy with the Total Differential Spectral Model (TDSM): Framework and Its Application to 11 Million LAMOST Low-Resolution Spectra }

\author{Lanya Mou}
\affiliation{National Astronomical Observatories, Chinese Academy of Sciences, Beijing 100101, China}
\affiliation{School of Astronomy and Space Science, University of Chinese Academy of Sciences, Beijing 100049, China}

\author{Maosheng Xiang}
\affiliation{National Astronomical Observatories, Chinese Academy of Sciences, Beijing 100101, China}
\affiliation{Institute for Frontiers in Astronomy and Astrophysics, Beijing Normal University, Beijing 102206, People’s Republic of China}

\author{Haibo Yuan}
\affiliation{Institute for Frontiers in Astronomy and Astrophysics, Beijing Normal University, Beijing 102206, People’s Republic of China}
\affiliation{School of Physics and Astronomy, Beijing Normal University, Beijing, 100875, China}

\author{Tianmin Wu}
\affiliation{National Astronomical Observatories, Chinese Academy of Sciences, Beijing 100101, China}
\affiliation{School of Astronomy and Space Science, University of Chinese Academy of Sciences, Beijing 100049, China}

\author{Meng Zhang}
\affiliation{National Astronomical Observatories, Chinese Academy of Sciences, Beijing 100101, China}
\affiliation{Max-Planck-Institut für Astronomie, Königstuhl 17, D-69117 Heidelberg, Germany}

\newcommand{\teff}{\ensuremath{T_{\mathrm {eff}}\,}}
\newcommand{\logg}{\ensuremath{{\mathrm {log}\, } g\,}}
\newcommand{\feh}{\ensuremath{[{\mathrm {Fe/H}}]\,}}
\newcommand{\afe}{\ensuremath{[{\mathrm {\alpha/Fe}}]\,}}
\newcommand{\mk}{\ensuremath{\mathrm M_K\,}}
\newcommand{\msun}{\ensuremath{M_\odot\,}}
\newcommand{\Hii}{\ensuremath{\mathrm {H\,\textsc{ii}}\,}}
\newcommand{\Ha}{\ensuremath{\mathrm H_\alpha\,}}
\newcommand{\age}{\ensuremath{\tau}}

\begin{abstract}
Deriving stellar astrophysical parameters from large-scale spectroscopic surveys requires accurate modeling of spectra across a wide wavelength coverage and a broad range of stellar parameter space, which remains a great challenge. In this work, we present \texttt{TDSM}, a physically motivated neural network architecture that self-consistently emulates both the flux and its first-order derivative of the ab initio stellar spectra. The method enables accurate and precise emulation of Kurucz synthetic spectra with severely blended features at $R \simeq 2000$ in a high-dimensional parameter space, achieving a typical pixel‑wise flux precision of 0.05\% in the wavelength range $3700<\lambda<9000${\AA} over the parameter range of $3000 < T_{\rm eff} < 45,000$~K, $-1 < \log g < 5$, and $-4.0 < \mathrm{[Fe/H]} < 0.5$. The model is applied to LAMOST DR10 to derive stellar parameters for 11 million spectra spanning the full model parameter range. 
We conduct detailed refinement and calibration efforts to mitigate systematic errors, which are revealed to be significant and complex due to imperfections in the ab initio model atmosphere. The resultant catalog provides eight stellar labels -- $T_{\rm eff}$, $\log g$, $v_{\rm mic}$, [Fe/H], [$\alpha$/Fe], [C/Fe], [N/Fe], and [Mg/Fe] -- for 8 million unique stars (11 million spectra including repeat observations) across the full parameter space. This adds a valuable dataset for studies of Galactic structure and archaeology, stellar populations, and related areas of astrophysics.
\end{abstract}


\section{Introduction} 
The accurate and precise determination of stellar parameters is essential for releasing the scientific potentials of large-scale spectroscopic surveys, such as the Large Sky Area Multi-Object Fiber Spectroscopic Telescope \citep[LAMOST;][]{2012RAA....12.1197C, 2022Innov...300224Y}, the Dark Energy Spectroscopic Instrument \citep[DESI;][]{2023ApJ...947...37C}, SDSS-V \citep{2017arXiv171103234K}, and the 4-metre Multi-Object Spectrograph Telescope \citep[4MOST;][]{2019Msngr.175....3D}. 
For low-resolution survey spectroscopy, this task is challenging in at least two aspects. First, it requires accurate spectral modeling or emulation across a high-dimensional parameter space to account for severe line-blending effects \citep{2017ApJ...843...32T}. Second, it demands reliable models that are valid over a vast range of parameters, accommodating the survey’s extensive coverage of the color–magnitude diagram \citep[e.g.][]{2014IAUS..298..310L, 2015MNRAS.448..855Y}.

Machine learning offers a practical approach for emulating stellar spectra across high-dimensional parameter spaces. It has been demonstrated that even at spectral resolutions of $R\lesssim2000$, machine learning models can effectively derive abundances for more than 20 chemical elements, and deliver atmospheric parameters with high precision, achieving internal uncertainties within 30~K in effective temperature, and $\simeq$0.05~dex in surface gravity and chemical abundances \citep[e.g.][]{2019ApJS..245...34X, 2020ApJ...898...58W, 2025ApJS..279....5Z}.

So far, the success of machine learning models in low-resolution spectroscopy has largely relied on their application within a data-driven framework. In this approach, spectral models are constructed self-consistently using a subset of survey data that have reference labels from high-resolution spectroscopy, ensuring high internal precision and minimal systematic errors in parameter inference. This approach, however, entails a fundamental compromise: the spectral emulation remains limited both in parameter dimensionality and in parameter space coverage, and hardly match the full scope of modern low-resolution surveys. Moreover, to accurately emulate spectral features for individual parameters, it is necessary to incorporate prior knowledge of their spectral response functions, as derived from $ab$ $initio$ modeling \citep{2017ApJ...849L...9T, 2019ApJS..245...34X}.

Model-driven approaches offer distinct advantages for stellar parameter determination, as they are built upon physically-motivated spectral modeling with minimal limitations in the dimensionality and coverage of the parameter space. While such methods are commonly employed to derive atmospheric parameters and elemental abundances from high-resolution spectra \citep[e.g.,][]{2016AJ....151..144G, 2019ApJ...879...69T, 2025PASA...42...51B}, their application to low-resolution survey spectra remains relatively limited.

Early algorithms and tools for model-driven analysis of low-resolution spectra relied on interpolating spectra from a pre-computed grid uniformly distributed in parameter space \citep[e.g.,][]{2008AJ....136.2022L, 2015AAS...22542207A, 2016RAA....16..110L}. This approach is computationally demanding and typically yields only a few basic stellar atmospheric parameters \citep{2008AJ....136.2022L, 2015AJ....150..187L, 2023ApJ...947...37C}.

\texttt{The Payne} \citep{2019ApJ...879...69T} pioneered the use of machine learning to emulate ab initio stellar spectra in a high-dimensional parameter space from a sparse, non-uniform grid of templates. This method has subsequently been adopted for low-resolution spectra. For instance, \citet{2022A&A...662A..66X} and \citet{2025A&A...698A.300S} applied it to determine atmospheric parameters and elemental abundances for hot stars ($T_{\rm eff}>7000$~K) from LAMOST low-resolution spectra ($R\simeq1800$). \citet{2025arXiv251106546V} demonstrated its application in deriving stellar abundances from 4MOST-like spectra ($R\simeq5000$). Beyond \texttt{The Payne}, \citet{2025ApJS..277...47Z} used a model-driven machine learning model with domain adaptation to determine parameters for M-type stars from LAMOST spectra, while \citet{2026arXiv260402730W} employed a model-driven deep learning approach to derive abundances for FGK-type stars from LAMOST low-resolution spectra.

Despite these advances, a model-driven pipeline capable of operating in high dimensions and uniformly across the broad parameter space representative of a survey's full stellar populations is still largely absent. In this work, we introduce the Total Differential Spectral Model (\texttt{TDSM}), a machine-learning method designed for accurately emulating both the flux and their first-order derivatives of the ab initio model stellar spectra. 
\texttt{TDSM} stands on the shoulders of \texttt{The Payne}, as it adopts a similar neural network architecture, but presents meaningful advance by explicitly emulates both the spectral flux and its first-order partial derivatives with respect to individual stellar parameters. Since these derivatives encode the true spectral features that are informative for parameter inference, their accurate emulation is crucial for ensuring rigorous and physically interpretable results.

We employ \texttt{TDSM} to determine stellar parameters from approximately 11 million low-resolution spectra in LAMOST DR10, covering a broad effective temperature range of 2,500~K to 45,000~K. For this application, \texttt{TDSM} is used to emulate spectra and derive eight key stellar parameters: effective temperature ($T_{\rm eff}$), surface gravity ($\log g$), microturbulent velocity ($v_{\rm micro}$), metallicity ([Fe/H]), $\alpha$-element enhancement ([$\alpha$/Fe]), and the abundance ratios [C/Fe], [N/Fe], and [Mg/Fe]. 

Not surprisingly, the inferred stellar parameters exhibit non-negligible systematic errors due to imperfections in the synthetic spectra. To mitigate this effect, the raw parameter estimates have been calibrated against reference data sets from literature. The final product is a catalog of stellar parameters derived from rigorous spectral fitting, achieving high internal consistency across a broad parameter space.

The structure of this paper is as follows. Section~\ref{sec:method} outlines the overall methodology, describing the theoretical training set, the construction of the differential spectra, and the neural network architecture. In Section~\ref{sec:Application to LAMOST DR10}, we turn to the application of the method to LAMOST DR10 spectra. Section~\ref{sec:calibration and validation} focuses on calibration and validation of the results. The resulting stellar parameter catalog for LAMOST DR10 is then presented in Section~\ref{sec:The stellar parameter catalog for LAMOST DR10}. Section~\ref{sec:Discussion} offers a discussion of the implications and remaining limitations of our method, followed by a summary of the main results.

\section{Method}\label{sec:method}
We determine stellar parameters with a transparent forward modelling framework. The method operates in two stages: first, we construct a forward spectra model with neural network method by training on a grid of synthetic spectra across a broad parameter space. Second, it fits the spectra model to observed spectra to derive multiple stellar parameters simultaneously. 

\subsection{The total differential spectral model}
For spectra emulation, the question is how to generate the spectrum for a star with arbitrary parameters $\vec{\theta}$, given a grid of reference spectra (or a training set) with parameters \{$\vec{\theta}_{\rm ref}$\}. In linear approximation, the spectrum for a star with parameters $\vec{\theta}$ can be inferred from a nearby reference point $\vec{\theta}_{\rm ref}$ via 
\begin{equation}
f(\lambda; \vec{\theta}) = f(\lambda; \vec{\theta}_{\rm ref}) + \nabla_{\vec{\theta}\vert\vec{\theta}_{\rm ref}}f(\lambda;\vec{\theta})(\vec{\theta}-\vec{\theta}_{\rm ref}),
\end{equation}
where $\nabla_{\vec{\theta}\vert\vec{\theta}_{\rm ref}}$ is the first derivative of the flux to the parameters at $\vec{\theta}_{\rm ref}$. 

To emulate $f(\lambda; \vec{\theta})$ accurately, one requires not only a grid of reference spectra (i.e., a training set) $f(\lambda; \vec{\theta}_{\rm ref})$, but also precise knowledge of the gradient $\nabla_{\vec{\theta}\vert\vec{\theta}_{\rm ref}}$. In most practical cases, however, the gradient is either unknown or not explicitly used as an input. Instead, a typical machine learning approach trains solely on $f(\lambda; \vec{\theta}_{\rm ref})$ and learns $\nabla_{\vec{\theta}\vert\vec{\theta}_{\rm ref}}$ implicitly.
Unfortunately, this can lead to imperfect gradient estimates \citep{2017ApJ...849L...9T, 2026arXiv260402730W}, which may be attributed to: i) substantial covariances among gradients of different parameters, and ii) the presence of observational noise. To mitigate these issues, one can incorporate additional constraints on $\nabla_{\vec{\theta}\vert\vec{\theta}_{\rm ref}}$ derived from ab initio model calculations when training a data-driven model \citep[e.g.][]{2017ApJ...849L...9T, 2019ApJS..245...34X}.

Here we propose the Total Differential Spectral Model (\texttt{TDSM}), which generalizes the idea of \citet{2017ApJ...849L...9T} in a model-driven framework by treating $\nabla_{\vec{\theta}\vert\vec{\theta}_{\rm ref}}$ as a known dataset alongside $f(\lambda; \vec{\theta}_{\rm ref})$. In this framework, the machine learning model is trained simultaneously on the reference spectra and their corresponding gradients. In our implementation, $f(\lambda; \vec{\theta}_{\rm ref})$ and $\nabla_{\vec{\theta}\vert\vec{\theta}_{\rm ref}}$ are computed ab initio using the Kurucz model atmosphere \citep{1970SAOSR.309.....K, 1993sssp.book.....K, 2014dapb.book...39K}. Our current model application to LAMOST spectra incorporates seven stellar parameters: $T_{\rm eff}$, $\log~g$, $v_{\rm micro}$, [Fe/H], [$\alpha$/Fe], [C/Fe], and [N/Fe].

\subsection{The Kurucz synthetic spectra}\label{KURUCZ}
To build a library of synthetic spectra, we first define a grid of model parameters spanning a broad range in each of the seven dimensions introduced above. The detailed parameter ranges of the grid are provided in Table~1. In order to construct a spectral library that reflects physically plausible stellar populations while maintaining wide parameter coverage, we sample the fundamental atmospheric parameters $T_{\rm eff}$, $\log g$, and [Fe/H] from the MIST isochrones \citep{2016ApJ...823..102C}. The remaining parameters are randomly drawn from their respective given ranges. It should be noted that white-dwarf atmosphere models are not included in the synthetic spectral library, so white dwarfs are outside the scope of the present work.
\begin{table}[htbp]
\centering
  \caption{Stellar atmospheric parameter space of the synthetic library.}
  \label{tab:table1}
\setlength{\tabcolsep}{3mm}
    \begin{tabular}{ccc}
\toprule
parameter            & range         &  grid   \\ \midrule
$T_{\rm eff}$   & $[2500,\,50000]\,\mathrm{K}$ & MIST \\
$\log g$        & $[-1.0,\,5.5]$              & MIST \\
{\rm [Fe/H]}    & $[-4.0,\,0.5]$              & MIST \\
$v_{\rm micro}$ & $[0,\,10]\,\mathrm{km\,s^{-1}}$ & uniform \\
$[\alpha/{\rm Fe}]$ & $[-0.4,\,0.8]$          & uniform \\
{\rm [C/Fe]}    & $[-1.5,\,2.0]$              & uniform \\
{\rm [N/Fe]}    & $[-1.5,\,2.0]$              & uniform \\
\bottomrule
    \end{tabular}
\end{table}

Ultimately, the library comprises 14,901 model spectra $f(\lambda; \vec{\theta}_{\rm ref})$, whose distribution in the $T_{\rm eff}$-$\log g$ diagram is shown in Figure~1. For each model in this grid, we compute a plane-parallel model atmosphere using \texttt{ATLAS12} \citep{2003IAUS..210P.A20C} and generate a synthetic spectrum by solving the radiative transfer equation with the \texttt{SYNTHE} \citep{2003IAUS..210P.A20C}. We adopt the solar abundance scale of \citet{Asplund2009}, and an updated version of the Kurucz line list \citep{Kurucz2017}. An extensive list of molecules are included in the calculation, including H$_2$, HD, CH, NH, OH, NaH, MgH, SiH, CaH, CrH, FeH, C$_2$, CN, CO, MgO, AlO, SiO, VO, TiO, and H$_2$O. The spectra are generated to have a resolution of $R=500,000$, and then are degraded to match that of the LAMOST line spread function. The spectra are then normalized following the procedure introduced in Section~3.1. 

\begin{figure}[ht!]
    \centering
    \includegraphics[width=0.9\linewidth]{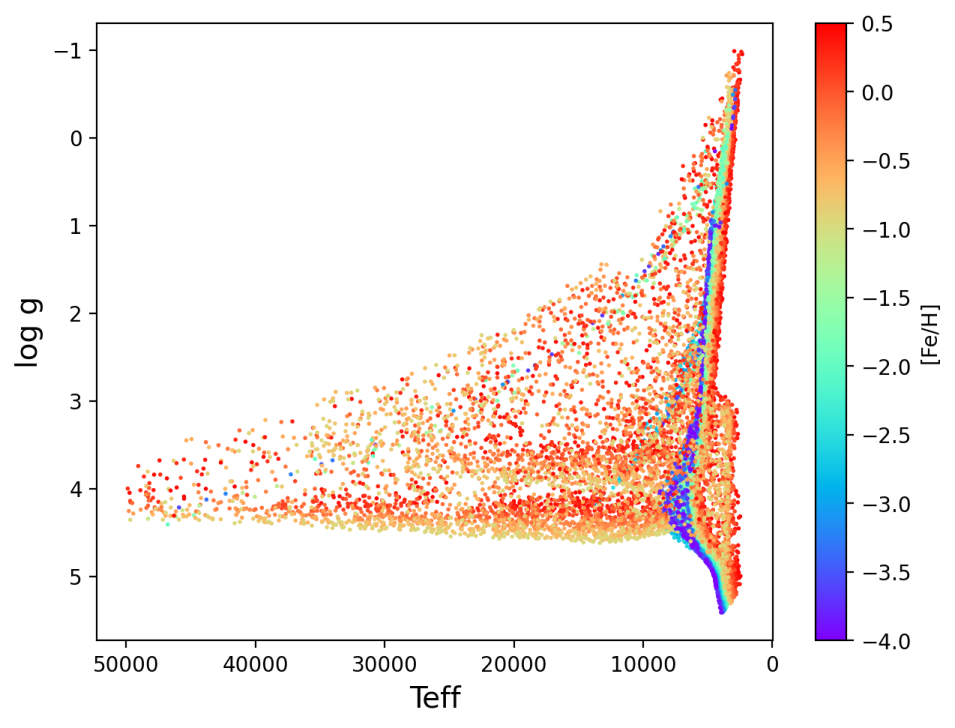}
    \caption{\label{fig:training_para_hr} 
Distribution of the Kurucz model atmosphere grid in the $T_{\rm eff}$-$\log~g$ diagram, color-coded by metallicity. The stellar parameters span 2500–50000~K in effective temperature ($T_{\rm eff}$), $-1$ to $6$ dex in surface gravity ($\log~g$), and $-4$ to $0.5$ dex in metallicity ([Fe/H]). }
\end{figure}

The differential spectra $\nabla_{\vec{\theta}\vert\vec{\theta}_{\rm ref}}$ for 689 randomly selected reference grids are computed self-consistently using \texttt{ATLAS12} and \texttt{SYNTHE}. In practice, we define the differential spectrum for a particular parameter $\theta_{\rm ref}$ as
\begin{equation}
\label{equ:differential_spectra}
\nabla_{\theta\vert\ \theta_{\rm ref}} := \frac{f(\lambda; \theta_{\rm ref}+\Delta{\theta}) - f(\lambda; \theta_{\rm ref})}{\Delta{\theta}}.
\end{equation}
We adopt a $\Delta{\theta}$ of 100~K in $T_{\rm eff}$, 0.2~dex in $\log~g$, and 0.2~dex in [Fe/H] and [X/Fe]. For each of the 689 reference grids, we generate seven differential spectra -- one for each stellar parameter.

\subsection{The neural network model design}
The model employs a fully connected neural network to perform forward modeling of stellar spectra. The model takes atmospheric parameters as inputs and predicts their corresponding spectra. Its architecture includes two hidden layers with 128 and 256 neurons, respectively, and is implemented using the \texttt{PyTorch} framework \citep{paszke2017automatic, paszke2019pytorch}. 

A defining feature of \texttt{TDSM} is that it trains the neural network model simultaneously on the flux $f(\lambda; \vec{\theta}_{\rm ref})$ and its gradients $\nabla_{\vec{\theta}\vert\vec{\theta}_{\rm ref}}$. This is implemented by adopting a tailored loss function as 

\begin{equation}
\begin{split}
\label{equ:regularization}
L
&= L_{\rm flux}+L_{\rm grad} \\
&=
\frac{w}{mn}
\sum_{k=1}^{m}\sum_{i=1}^{n}
\left|
f(\lambda_i;\boldsymbol{\theta}_k)
-
f_{\rm ground\,truth}(\lambda_i;\boldsymbol{\theta}_k)
\right| \\
&\quad +
\sum_{j}
\frac{w_j}{ln}
\sum_{k=1}^{l}\sum_{i=1}^{n}
\Bigl|
\nabla_{\theta_j} f(\lambda_i;\boldsymbol{\theta}_k) \\
&\qquad\qquad\qquad
-\nabla_{\theta_j} f_{\rm ground\,truth}(\lambda_i;\boldsymbol{\theta}_k)
\Bigr|.
\end{split}
\end{equation}

where $m$ is the number of training spectra, $n$ is the number of wavelength pixels in each spectrum, and $l$ is the number of spectral gradients adopted in the model. 
The vector $\boldsymbol{\theta}_k$ denotes the $k$th label set. 
The quantities $f(\lambda_i;\boldsymbol{\theta}_k)$ and $\nabla_{\theta_j} f(\lambda_i;\boldsymbol{\theta}_k)$ represent the predicted flux and predicted gradient respectively, while $f_{\rm  ground\,truth}$ and $\nabla_{\theta_j} f_{\rm  ground\,truth}$ are the corresponding spectrum and gradient spectrum from the Kurucz model. 
The first term, $L_{\rm flux}$, measures the mean flux difference averaged over all spectra and wavelength pixels with weights $w$. 
The second term, $L_{\rm grad}$, measures the gradient mismatch for each label, averaged over all spectra and pixels, and then summed over all $l$ labels with weights $w_j$.

For the training process, a total of 11,000 of the reference spectra $f(\lambda; \vec{\theta}_{\rm ref})$, and all the differential spectra $\nabla_{\vec{\theta}\vert\vec{\theta}_{\rm ref}}$, are adopted as the training set, while the remaining reference spectra $f(\lambda; \vec{\theta}_{\rm ref})$ serve as the validation set. The spectral model is trained by dividing the wavelength range into six segments. A separate neural network is trained for each segment. For each model, we train 30,000 epochs using a batch size of 256, and the learning rate is gradually reduced from 0.01 to 0.0001.

\subsection{Performance of the model from self validation}

\begin{figure*}[htp!]
    \centering
    \includegraphics[width=0.95\linewidth]{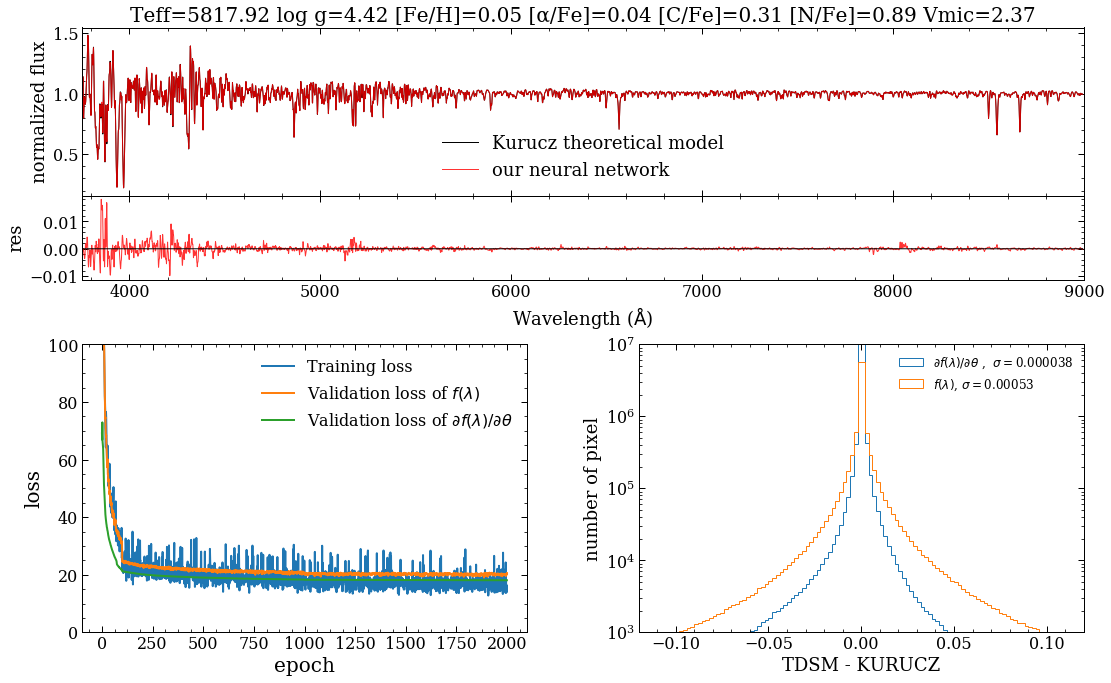}
    \caption{\label{fig:example+loss} An overview of the \texttt{TDSM} model performance. The top panels show the \texttt{TDSM} spectrum emulation of a solar-like star, compared to the Kurucz model computation. The bottom-left panel shows the \texttt{TDSM} training loss and validation loss of both the flux values and their gradients, for a wavelength pixel as an example. The bottom-right panel presents the residual distribution of validation Kurucz spectra sample and the gradients. The overall scatter is $\sigma = 0.00053$ in flux residuals, and $\sigma = 0.000038$ in the gradients.} 
\end{figure*}

Figure~\ref{fig:example+loss} shows an example of the emulated spectrum by our model, compared to the Kurucz ab initio model spectrum for a solar-like star. The residuals between the two spectra are smaller than $1\times10^{-3}$ for most of the pixels.
The bottom-left panel of figure~\ref{fig:example+loss} presents an example of how the loss decreases with training epoch for one wavelength pixel. All loss components, including the loss of the reference spectra, the differential spectra, and the loss of validation spectra, are well converged. The training loss is comparable to the validation loss, indicating little overfitting.
The bottom-right panel displays the pixel-wise flux-residual distribution between the TDSM emulation and the ab initio Kurucz model, derived from a validation dataset of 3,901 spectra. These residuals have a standard deviation of $5.3\times10^{-4}$. The same panel also shows the corresponding residual distribution for the flux gradients, evaluated across the full training dataset. This gradient residual exhibits a standard deviation of only $3.8\times10^{-5}$ -- more than an order of magnitude smaller than that of the flux residuals. Such low values highlight the exceptional internal precision achieved by the TDSM in reproducing the ab initio Kurucz spectra.
 
\begin{figure*}[htp!]
    \centering
    \includegraphics[width=0.9\linewidth]{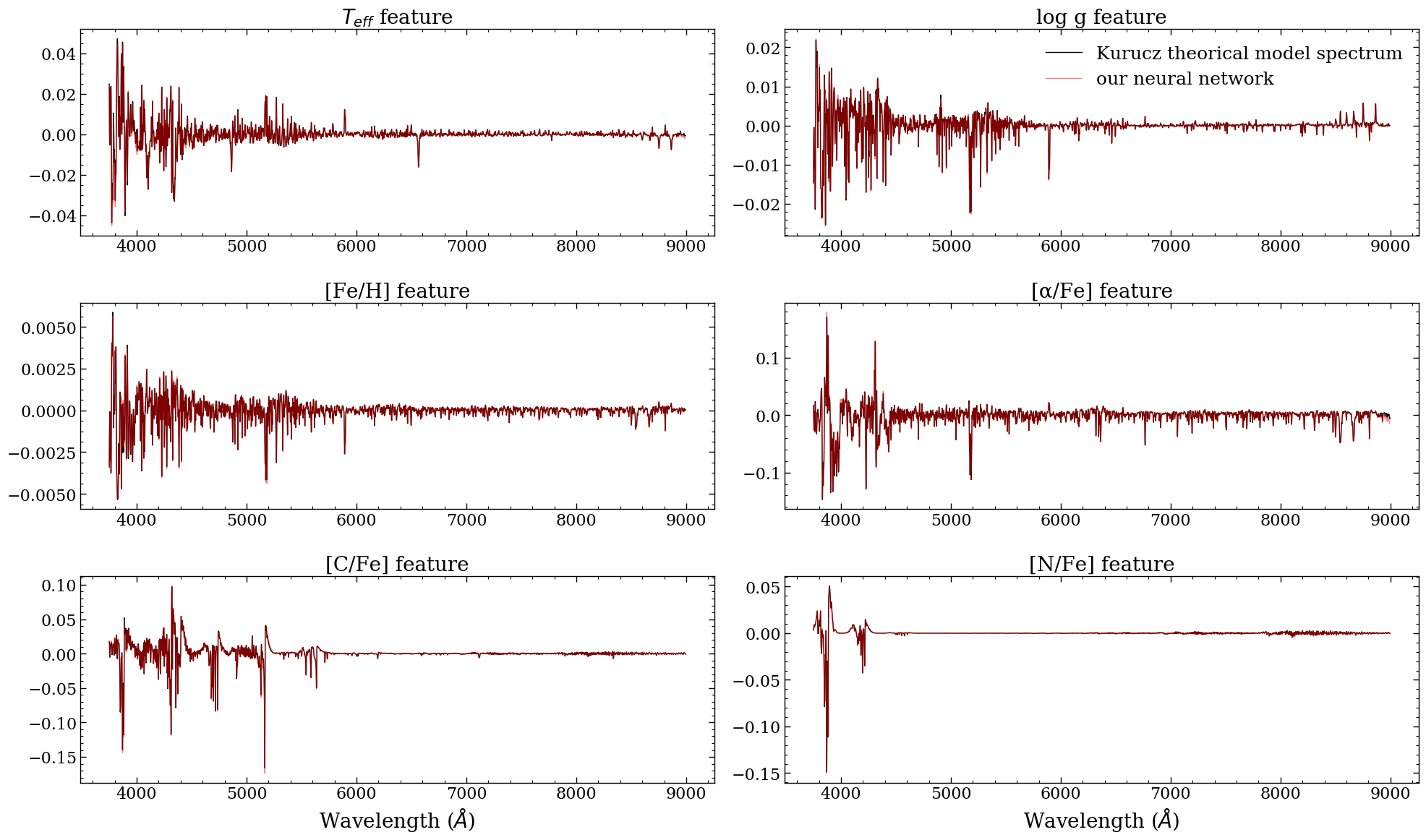}
    \caption{\label{fig:labels_features} Example of gradient spectra for different labels for a star with $T_{\rm eff}=5600\,\mathrm{K}$, $\log g = 4.0$, $\mathrm{[Fe/H]}=0$, $[\alpha/\mathrm{Fe}]=0$, $\mathrm{[C/Fe]}=0$, $\mathrm{[N/Fe]}=0$, and $V_{\rm mic}=1.16\,\mathrm{km\,s^{-1}}$. The black and red lines represent the gradients of the Kurucz model computation and the \texttt{TDSM} reconstruction, respectively.}
\end{figure*}

\begin{figure*}[htp!]
    \centering
    \includegraphics[width=0.9\linewidth]{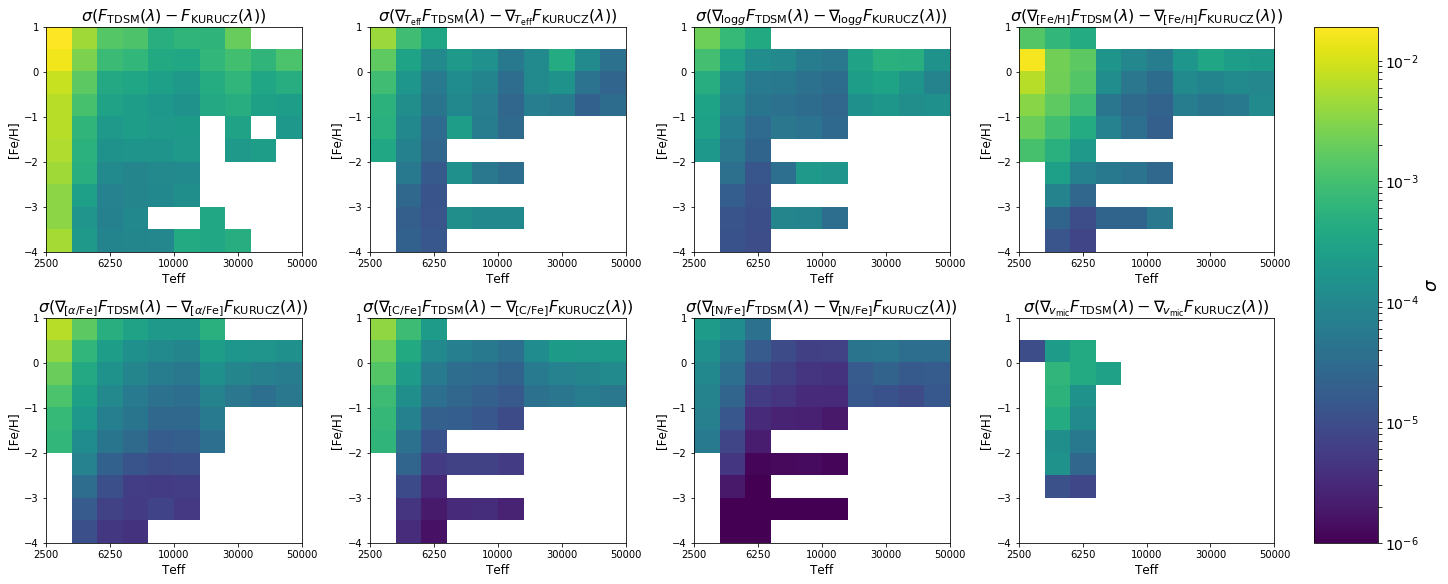}
    \caption{\label{fig:res_bins} Distribution of scatter in the residuals of the flux and gradients across the $T_{\rm eff}$–[Fe/H] plane. The first panel shows the results for the flux, while the remaining panels display the results for the gradients for different labels. The color scale represents the mean value of the standard deviation ($\sigma$) in the residuals for validation spectra in each $T_{\rm eff}$–[Fe/H] bin.} 
\end{figure*}

Figure~\ref{fig:labels_features} presents an example of the differential spectra generated by the \texttt{TDSM}. The model accurately reproduces the ab initio differential spectra for all parameters. For each of the gradient spectrum, we compute the standard deviation (dispersion) of the residuals between \texttt{TDSM} generation and the Kurucz ab initio computation for both the spectra and the differential spectra. Figure~\ref{fig:res_bins} shows the typical dispersion for stars of different $T_{\rm eff}$ and [Fe/H]. It shows that the dispersion, which indicates the errors of the spectra emulation, are larger for cooler or more metal-rich stars. Typical dispersion for FGK stars are smaller than $10^{-3}$ for the spectral flux, and smaller than $10^{-4}$ for the differential spectra.     

We apply the \texttt{TDSM} model to a set of test spectra in order to assess the precision of the derived stellar atmospheric parameters. The test set comprises 1309 FGK stellar spectra generated from the Kurucz model, with realistic noise added to mimic the characteristics of LAMOST spectra. To this end, we construct an empirical error model of the LAMOST spectra that includes two components: a Gaussian part accounting for random errors, and a non‑Gaussian part representing systematic errors introduced during observations (e.g., scattered light) and data reduction (e.g., fiber flat‑fielding, sky subtraction). The non‑Gaussian component is derived from the residuals of fits to real LAMOST spectra (see Appendix~\ref{sec:residual_model} for details). For simplicity, we restrict the current analysis to a constant S/N of 50; the results can be extrapolated to other S/N values using the parameter error versus S/N relation obtained from repeat observations, which indicates that the errors are approximately inversely proportional to S/N.

\begin{figure*}[htp!]
    \centering
    \includegraphics[width=0.9\linewidth]{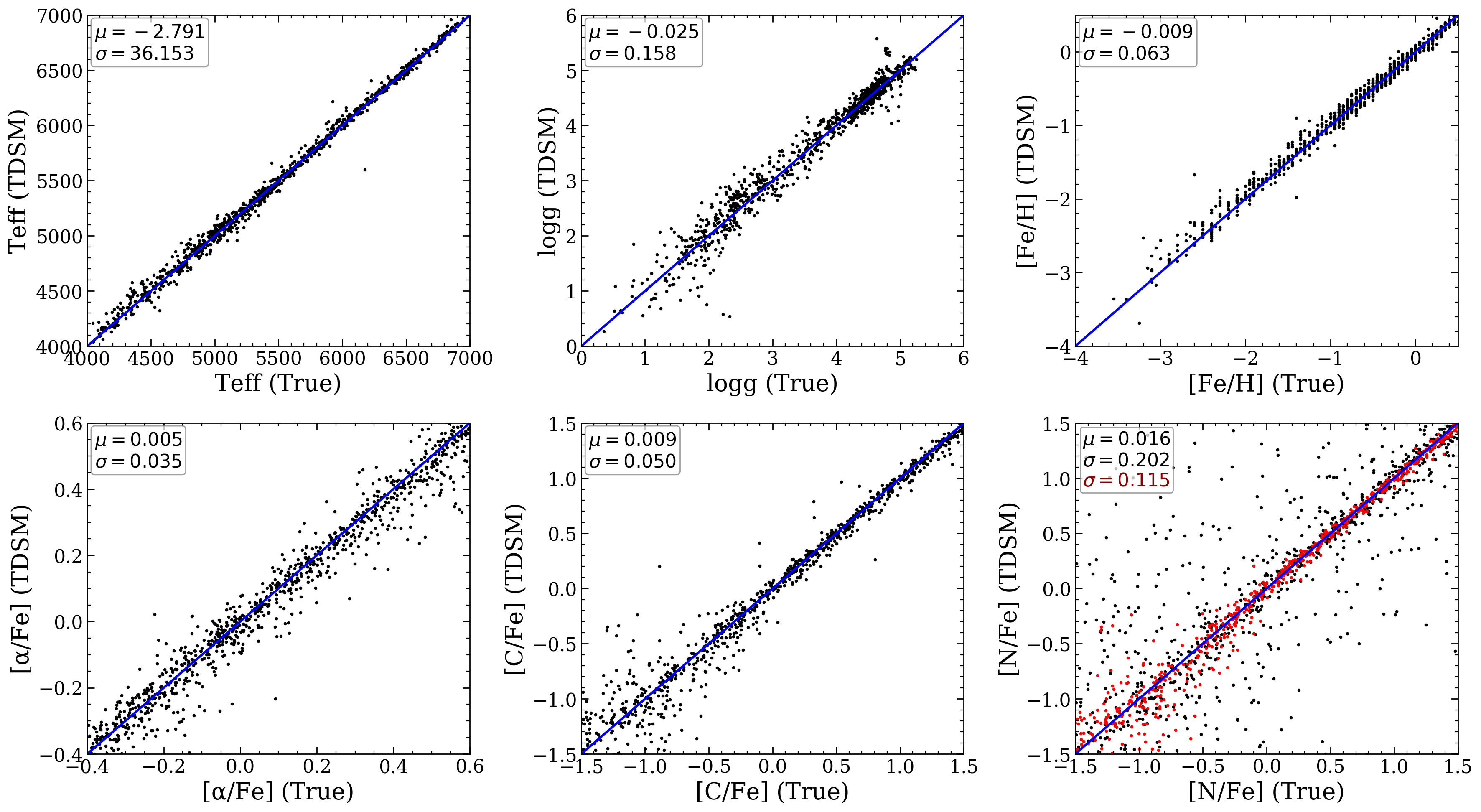}
    \caption{\label{fig:self_vali}  Self validation of the stellar-parameter determinations using synthetic test spectra. The horizontal axes show the true input parameters used in the Kurucz model calculation and spectral synthesis, while the vertical axes show the parameters recovered by \texttt{TDSM} fitting. 
    Mean and standard deviation of the residuals are shown in the upper-left corner of each panel. For [N/Fe], metal-rich stars with \({\rm [Fe/H]}>-1\) are highlighted with red dots, and the standard deviation for both the overall sample and the metal-rich sample are presented. }
\end{figure*}

We derive the stellar parameters from the test spectra using the {\sc Curvefit} Python package, which employs the TDSM spectral model and the Levenberg–Marquardt algorithm. Figure~\ref{fig:self_vali} compares the best‑fit parameters with the ground‑truth values used to generate the Kurucz spectra. The standard deviations of the residuals for the whole sample are $\sim 36$~K in $T_{\mathrm{eff}}$, $\sim 0.16$~dex in $\log g$, and $\sim 0.06$~dex in [Fe/H], 0.04~dex in [$\alpha$/Fe], 0.05~dex in [C/Fe], and 0.20~dex in [N/Fe]. The scatter in $\log~g$ depends on $\log~g$ itself, as dwarfs ($\log~g>4$) show a standard deviation of only 0.11~dex, whereas giants yield 0.23~dex. These values are consistent with independent checks based on repeat LAMOST observations (Section 4). The uncertainty in [N/Fe] is strongly metallicity‑dependent, because the nitrogen features (mainly the CN $\lambda$4227 band) weaken rapidly as metallicity decreases. For metal‑rich stars with ${\rm [Fe/H]}>-1$, the standard deviation drops to only 0.12~dex.

The above standard deviations provide an estimate of the parameter precision achievable with the current method for LAMOST‑like spectra at S/N = 50. Nevertheless, we caution that because the test set consists of Kurucz model spectra rather than real observed data, the examination shown in Figure~\ref{fig:self_vali} does not account for systematic errors that may affect parameter estimates for actual LAMOST spectra.

\section{Application to LAMOST DR10} \label{sec:Application to LAMOST DR10}
\subsection{The LAMOST DR10 spectra}

We apply the spectral model to determine stellar parameters from the low-resolution spectra of LAMOST DR10\footnote{\url{http://www.lamost.org/dr10/}}.
LAMOST DR10 released $11,931,197$ spectra with medium spectral resolution ($R\simeq7500$; \citealt{2020arXiv200507210L}) and $11,817,425$ spectra with low spectral resolution ($R\lesssim1800$), together with their classifications, redshifts (radial velocities), and basic stellar parameters determined by the LAMOST pipelines \citep{2015RAA....15.1095L}.
In this work, we focus on the analysis of the LAMOST low-resolution spectra. While the LAMOST survey collects low-resolution spectra for both Galactic and extragalactic objects \citep{2012RAA....12..723Z}, stars occupy the vast majority of survey spectra data set as they are bright and numerous enough for the survey strategy. In LAMOST DR10, there are 11,473,640 spectra out of the 11,817,425 total spectra data set are classified as `stars'. These spectra are collected during Oct. 2011 and June, 2022 as a result of several survey campaigns \citep{2012RAA....12..735D, 2014IAUS..298..310L, 2015MNRAS.448..855Y}.

The spectra are pre-processed in several steps. First, we shift the spectra to rest frame by linear interpolation using line-of-sight velocity ($V_{\rm los}$) derived with the LAMOST pipeline. Then we normalize the spectra by dividing a smoothed version of themselves,
\begin{equation}
     f_{\rm n}(\lambda) = \frac{f(\lambda)}{\bar{f}(\lambda)}.
\end{equation}
where the smoothed spectra ${\bar{f}(\lambda)}$ are derived by convolving a Gaussian kernel via,
\begin{equation}
  \bar{f}(\lambda_i) = \frac{\Sigma_jf_j\omega_j(\lambda_i)}{\Sigma_j\omega_j(\lambda_i)},
\end{equation}
\begin{equation}
  \omega_j(\lambda_i) = \frac{1}{\sqrt{2\pi}L}\exp\left(-\frac{(\lambda_j-\lambda_i)^2}{2*L^2}\right).
\end{equation}

It was common to adopt a uniform value of kernel width $L$ for the spectra normalization \citep[e.g.][]{2017ApJ...836....5H, 2019ApJS..245...34X}. While in practice this has been a good choice for data-driven spectroscopic analysis for typical FGK stars, we find that it is an oversimplification for robust stellar parameter determination across the full parameter space, from hot, OBA stars with clean features in their spectra, to very cold, M stars with strong features dominated by molecular bands. Instead, for each LAMOST spectrum, we produce two normalized spectra using empirically selected kernel widths of 25~{\AA} and 100~{\AA}, respectively.
These two value are determined based on empirical test, by trying different values like 10, 25, 50, 100, 200, 400.
The narrow kernel (25~{\AA}) provides good normalization for atomic lines, while spectra normalized using the wider kernel (100~{\AA}) retains broader molecular bands in low-temperature ($T_{\rm eff}\lesssim4500$~K) stars.

As a result, two sets of stellar parameters are derived for each star using these  spectra normalized with different kernels. The choice and combination of the resultant parameters are discussed in Section~{3.2}. 

Finally, we mask bad pixels in the spectra by setting extremely large values in their flux errors. The bad pixels include those flagged by the 'andmask', 'ormask' labels of the LAMOST pipeline, and those have unrealistic (zero) values of flux or invariance.

\subsection{Stellar parameter fitting}
\begin{figure*}[htp!]
    \centering
    \includegraphics[width=0.9\linewidth]{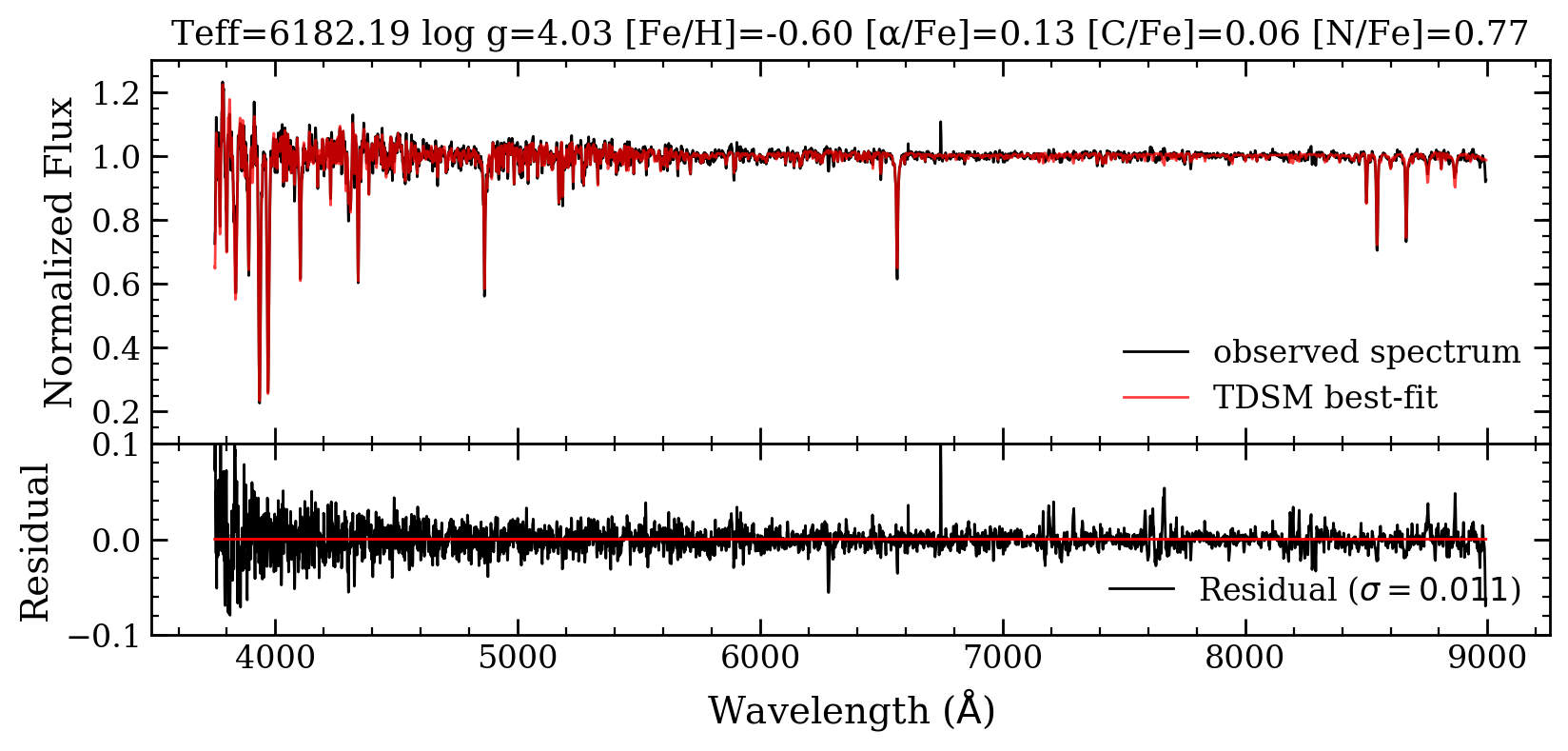}
    \caption{\label{fig:fit_sample} An example application of the LAMOST spectrum fitting. The normalized observed spectrum with $SNR_g=64.55$ is fitted to derive stellar parameters.  The residuals between the normalized observed spectrum and the best-fit spectrum are shown in the bottom panel. }
\end{figure*}

The Levenberg-Marquardt algorithm \citep{levenberg1944method, marquardt1963algorithm} in the {\sc Curvefit} Python package is adopted to fit stellar parameters by matching the spectral model to LAMOST spectra. Figure \ref{fig:fit_sample} shows an example application of the curve-fitting method, where the normalized LAMOST spectrum is fitted to derive stellar parameters. The residuals between the normalized LAMOST spectrum and the best-fit model spectrum are shown in the bottom panel. The scatter of the residuals is 0.011, which is contributed by both observational noise and systematic differences between synthetic and observed spectra.

To do so, there are several issues that need dedicated efforts: i) the combination for results from spectra normalized with different kernel widths, as introduced above; ii) the fitting results show dependence on the initial guess of stellar parameters; and iii) the presence of an artificial desert of A-type stars as a consequence of the inflection nature in their gradient spectra.  

\subsubsection{results for spectra normalized with different kernels}
The selection of an appropriate Gaussian kernel width for spectral normalization is a heuristic exercise, guided by inspection and iterative refinement. On one hand, the kernel width should be not small enough to retain the power for removing broad patterns that are not intrinsic features of the stellar SEDs, such as interstellar extinction that varies star-by-star, and potential calibration errors and artifacts in the data reduction. On the other hand, the kernel width should be large enough to retain as much line absorption features as possible in the normalized spectra.  

We have adopted two sets of kernel widths, namely 25~{\AA} and 100~{\AA}. The former is found to be an appropriate value for the normalization of relatively warm stars whose spectra are dominated by atomic lines, while the latter is a better value for the normalization for cold, M-type stars whose spectra are dominated by broad molecular bands.   

Each of these two kernels are applied to the normalization for all the LAMOST DR10 spectra, as well as subsequent stellar parameter fitting. This leads to two sets of parameter estimates for each of the spectrum, $\boldsymbol{\theta}_{\rm k25}$ and $\boldsymbol{\theta}_{\rm k100}$, respectively. These two sets of parameters are then combined into a single set by selecting the one with better fit for each spectrum based on both the $\chi^2$ of the spectral fitting and the correlation coefficient between the observed and best-fit spectrum, via the criteria
\begin{equation}
   \boldsymbol{\theta}_{\rm adopted}  = 
    \begin{cases} 
   \boldsymbol{\theta}_{\rm k100}, & \text{if } \chi^2_{\rm k25} - \chi^2_{\rm k100} > 1, \\
      & \text{or } \frac{R_{\rm k100}-R_{\rm k25}}{\sigma({R_{k100}-R_{\rm k25}})} > 5, \\
    \boldsymbol{\theta}_{\rm k25}, & \text{otherwise}.
    \end{cases} 
\end{equation}

Here $\chi^2_{\rm k25}$ and $\chi^2_{\rm k100}$ are the $\chi^2$ values of the best fit to normalized spectra with the sets of kernel widths, respectively. $R_{\rm k25}$ is the correlation coefficient between the best-fit model spectrum and the LAMOST spectrum normalized with a kernel width of 25~{\AA}, dubbed as $f_{n,\,\rm k25}(\lambda)$. While $R_{\rm k100}$, however, is the correlation coefficient between $f_{n,\,\rm k25}(\lambda)$ and the model spectrum calculated for the parameter set $\boldsymbol{\theta}_{\rm k100}$. $\sigma({R_{k100}-R_{\rm k25}})$ refers the standard deviation between the two sets of correlation coefficients. 

The [Mg/Fe] estimation is an exception. The spectra fitting derive the seven intrinsic parameters of the model, namely, $T_{\rm eff}$, $\log~g$, $v_{\rm micro}$, [Fe/H], [$\alpha$/Fe], [C/Fe], and [N/Fe]. After deriving these parameters, the parameters other than [$\alpha$/Fe] are fixed to re-fit the [Mg/Fe] from spectra regimes that are sensitive to Mg abundances using a uniform set of line mask. For simplicity, [Mg/Fe] derived from spectra normalized with the kernel width of 25~{\AA} are adopted for all the stars. 

\subsubsection{Initial value for parameter fitting}

We find that the results of parameter fitting are strongly influenced by the choice of initial values. This behavior is mainly caused by a sign change in the spectral gradients around A0-type stars, where the Balmer lines reach their maximum strength. When a low effective temperature, for example the solar value, is adopted as the initial value, intrinsically hot stars tend to converge to a cooler local minimum, yielding artificially low temperature estimates. This occurs because the small spectral gradients near the A0-type regime act as an inflection region in the optimization process. Conversely, when hot-star parameters are used as the initial values, some non-hot stars may be incorrectly driven toward the hot-star solution.

To deal with this issue, we perform the parameter fitting twice using two sets of initial values, namely 
$\text{set}_1 = \{T_{\rm eff}=5800\,{\rm K},\ \log g = 4.5,\ [{\rm Fe/H}] = 0,\ v_{\rm mic}=1\}$ 
and 
$\text{set}_2 = \{T_{\rm eff}=15000\,{\rm K},\ \log g = 4.5,\ [{\rm Fe/H}] = 0,\ v_{\rm mic}=2\}$.
All the abundance ratios are set to be 0.
The two solutions are then compared and combined to a single set of final result, using the following criterion, 

\begin{equation}
   \boldsymbol{\theta}_{\rm adopted} =
   \begin{cases} 
      \boldsymbol{\theta}_{\rm ini\text{-}hot\ star}, &
      \begin{aligned}
         & C_1 \lor C_2 \lor C_3 \lor C_4 \\
         & \lor C_5 \lor C_6 \lor C_7 ,
      \end{aligned} \\[6pt]
      \boldsymbol{\theta}_{\rm ini\text{-}sun}, & \text{otherwise}, 
   \end{cases}
   \label{equ:hot_star}
\end{equation}
where $\lor$ denotes ''or''. The $C_i$ ($i=1,\dots,7$) represent the criteria used to identify hot stars (Equations~\ref{equ:hot_c1}--\ref{equ:hot_c7})

The identification of intrinsic hot stars is a complex task. Hot stars with spectral types O or B exhibit prominent He\,{\sc i} ($\lambda 4471$\,\AA) and He\,{\sc ii} ($\lambda 4686$\,\AA) lines, but relatively weak Ca\,{\sc ii} K ($\lambda 3933$\,\AA) lines. The equivalent widths (EW) of these lines (as shown in Equation~\ref{equ:ew}) therefore provide direct indicators of their nature. 

\begin{equation}
    \mathrm{EW} = \int_{\lambda_1}^{\lambda_2} \bigl(1 - f_\lambda\bigr)\, d\lambda
    \label{equ:ew}
\end{equation}
and its uncertainty is estimated by standard error propagation assuming uncorrelated flux errors
\begin{equation}
    \sigma_{\mathrm{EW}} = \left[ \sum_{i} \left(\Delta \lambda_i \, \sigma_{f,i}\right)^2 \right]^{1/2}
    \label{equ:ew_err}
\end{equation}
where $\mathrm{EW}$ denotes the equivalent width measured over the wavelength interval $[\lambda_1, \lambda_2]$. 
The quantity $f_\lambda$ represents the normalized flux at wavelength $\lambda$, and $f_i$ is its discrete counterpart at pixel $i$. 
The term $\Delta \lambda_i$ is the wavelength interval associated with the $i$-th pixel. 

However, for spectra with low S/N, the EW of these lines may suffer from large uncertainties to allow robust identification. Therefore, additional diagnostics are required to achieve precise classification. In this work, we also compute $\chi^2$ values of the fits for wavelength windows around these lines. Furthermore,  we take prior from the spectral classifications provided by the LAMOST pipeline \citep{2015RAA....15.1095L}. The ultimate criteria of hot stars are as follows:

\begin{align}
C_1 :=\;&
\begin{aligned}[t]
& (\chi^2_{\mathrm{CaK},\,\mathrm{ini\text{-}sun}} \ge \chi^2_{\mathrm{CaK},\,\mathrm{ini\text{-}hot}}) \\
& \land (\mathrm{corr}_{\text{ini-sun}} \le \mathrm{corr}_{\text{ini-hot}}) \\
& \land \left(\dfrac{\mathrm{EW}_{\mathrm{CaK}} / \sigma_{\mathrm{EW}(\mathrm{CaK})}}
      {\mathrm{EW}_{\mathrm{CaH}} / \sigma_{\mathrm{EW}(\mathrm{CaH})}} \le 0.4\right)  \\
& \land (T_{\rm eff} > 5000\,\mathrm{K})
\end{aligned}
\label{equ:hot_c1}
\\[6pt]
C_2 :=\;&
\begin{aligned}[t]
& (\chi^2_{\mathrm{He\,I},\,\mathrm{ini\text{-}sun}} \ge \chi^2_{\mathrm{He\,I},\,\mathrm{ini\text{-}hot}}) \\
& \land (\chi^2_{\mathrm{CaK},\,\mathrm{ini\text{-}sun}} + 0.5 \ge \chi^2_{\mathrm{CaK},\,\mathrm{ini\text{-}hot}}) \\
& \land (T_{\rm eff} > 7000\,\mathrm{K})
\end{aligned}
\label{equ:hot_c2}
\\[6pt]
C_3 :=\;&
\begin{aligned}[t]
& (5500 < T_{\rm eff} < 7500\,\mathrm{K}) \land
  (T_{\rm eff}^{\mathrm{hot}} > 12000\,\mathrm{K}) \\
& \land (\tfrac{\mathrm{EW}_{\mathrm{He\,I}}}{\sigma_{\mathrm{EW}(\mathrm{He\,I})}} \ge 3 \lor \tfrac{\mathrm{EW}_{\mathrm{He\,II}}}{\sigma_{\mathrm{EW}(\mathrm{He\,II})}} \ge 3) \\
& \land (\tfrac{\mathrm{EW}_{\mathrm{CaH}}}{\sigma_{\mathrm{EW}(\mathrm{CaH})}} - 3\,\tfrac{\mathrm{EW}_{\mathrm{CaK}}}{\sigma_{\mathrm{EW}(\mathrm{CaK})}} > 15) \\
& \land (\tfrac{\mathrm{EW}_{\mathrm{CaH}}}{\sigma_{\mathrm{EW}(\mathrm{CaH})}} > 10) \land
  (\tfrac{\mathrm{EW}_{\mathrm{CaK}}}{\sigma_{\mathrm{EW}(\mathrm{CaK})}} < 10) \\
& \land (\sigma_{\mathrm{EW}(\mathrm{CaH})} < 100) \land
  (\sigma_{\mathrm{EW}(\mathrm{CaK})} < 100) \\
& \land (\sigma_{\mathrm{EW}(\mathrm{He\,I})} < 100) \land
  ([\mathrm{C/Fe}] \le 1.8)
\end{aligned}
\label{equ:hot_c3}
\\[6pt]
C_4 :=\;&
\begin{aligned}[t]
& (5500 < T_{\rm eff} < 6000\,\mathrm{K}) \land
  (T_{\rm eff}^{\text{hot}} > 20000\,\mathrm{K}) \\
& \land (\tfrac{\mathrm{EW}_{\mathrm{He\,I}}}{\sigma_{\mathrm{EW}(\mathrm{He\,I})}} \ge 3) \\
& \land (\sigma_{\mathrm{EW}(\mathrm{He\,II})} < 100) \\
& \land (\tfrac{\mathrm{EW}_{\mathrm{CaK}}}{\sigma_{\mathrm{EW}(\mathrm{CaK})}} < 3)
\end{aligned}
\label{equ:hot_c4}
\\[6pt]
C_5 :=\;&
\begin{aligned}[t]
& \bigl(\mathrm{subclass} \in \{\mathrm{B9}, \mathrm{B6}, \mathrm{O}, \mathrm{OB}\}\bigr)
\end{aligned}
\label{equ:hot_c5}
\end{align}


\begin{align}
C_6 :=\;&
\begin{aligned}[t]
& (T_{\rm eff} > 5800\,\mathrm{K}) \land
  \left(\frac{\mathrm{EW}_{\mathrm{He\,I}}}{\sigma_{\mathrm{EW}(\mathrm{He\,I})}} > 5\right) \\
& \land \left(\chi^2_{\mathrm{CaK},\,\mathrm{ini\text{-}hot}}
      \le \chi^2_{\mathrm{CaK},\,\mathrm{ini\text{-}sun}}\right) \\
& \land \left(
\dfrac{\mathrm{EW}_{\mathrm{CaK}} / \sigma_{\mathrm{EW}(\mathrm{CaK})}}
{\mathrm{EW}_{\mathrm{CaH}} / \sigma_{\mathrm{EW}(\mathrm{CaH})}}
< 0.3\right)
\end{aligned}
\label{equ:hot_c6}
\\[6pt]
C_7 :=\;&
\begin{aligned}[t]
& \left(\left|\frac{\mathrm{EW}_{\mathrm{CaK}}}
{\sigma_{\mathrm{EW}(\mathrm{CaK})}}\right| < 5\right) \\
& \land \left(T_{\rm eff}^{\mathrm{hot}} > 12000\,\mathrm{K}\right) \\
& \land (T_{\rm eff} > 6000\,\mathrm{K}) \\
& \land \left(\chi^2_{\mathrm{He\,I},\,\mathrm{ini\text{-}sun}}
      \ge \chi^2_{\mathrm{He\,I},\,\mathrm{ini\text{-}hot}}\right) \\
& \land (\mathrm{SNR}_2 > 20)
\end{aligned}
\label{equ:hot_c7}
\end{align}

with these criteria, we are able to maximally identify hot stars, not only those with high spectral S/N but also those  with poor spectral quality.

\subsubsection{Parameter fitting scheme for A-type stars}
We find the above parameter estimates exhibit an artificial desert for early A-type stars, in the temperature window of 9000--10000\,K. Similar issue has also been found in previous work (\citealt{2022A&A...662A..66X}). This is likely because, as mentioned above, the spectral gradients of the Balmer lines reach a minimum for A-type stars and act as an inflection region, leading to artificial bias due to the scatter of temperature estimates on both sides.

To mitigate this artifact, we adopt a similar approach to that of \citet{2022A&A...662A..66X}. We determine the parameters of early A-type stars only using spectra in which the Balmer lines are masked. In practice, this is implemented in several steps. First, we derive a set of parameters, denoted by $\boldsymbol{\theta}_{\rm mask_H}$, using spectra with masked Balmer lines for all the LAMOST sample stars. In this set of parameters, the A-type desert is effectively removed. However, the resulting parameters for either cooler or hotter of stars show degraded accuracy. We therefore combine the two set of parameters using the weighted scheme using the following equation defined in Equation~\ref{equ:weight}

\begin{equation}
\boldsymbol{\theta}_{\rm adopted} =
\begin{cases}
\begin{aligned}
\displaystyle
\frac{\boldsymbol{\theta}_{\rm mask_H}\, w_1
+ \boldsymbol{\theta}_{\rm allspec}\, w_2}{w_1 + w_2}, \\
\text{if } 8000 < T_{\rm eff}(\boldsymbol{\theta}_{\rm mask_H}) < 12000 \\
\text{and } T_{\rm eff}(\boldsymbol{\theta}_{\rm allspec}) > 6800
\end{aligned}
\\[10pt]
\boldsymbol{\theta}_{\rm allspec}, 
\quad \text{otherwise.}
\end{cases}
\label{equ:weight}
\end{equation}

where $\boldsymbol{\theta}_{\rm allspec}$ represents the set of parameters derived as described above from the full spectra. 
We adopt a Hann function shown in Equation~\ref{equ:Hann} to compute the weights for $\boldsymbol{\theta}_{\rm mask_H}$ and $\boldsymbol{\theta}_{\rm allspec}$.
\begin{equation}
    \omega_1(x) = 
    \begin{cases} 
    \cos^2 \left( \frac{\pi x}{L} \right), & \text{if } |x| \leq \frac{L}{2}, \\ 
    0, & \text{if } |x| > \frac{L}{2}.
    \end{cases}
    \label{equ:Hann}
\end{equation}

where $x = T_{\rm eff}(\boldsymbol{\theta}_{\rm allspec}) - 9500 K$. L for $T_{\rm eff}$ lower than 9500K is adopted as 3000,  L for $T_{\rm eff}$ greater than 9500K is adopted as 5000. $\omega_2 = 1 - \omega_1$. 

\subsection{Data Flag}\label{data_flag}

To quantify the quality of spectral fitting, we define two quality flags in the final catalog: a global indicator, ${\rm flag}_{{\chi}^2}$ and line-based indicators, ${\rm flag}_{R_{\rm line}}$.

We first define a ${\rm flag}_{{\chi}^2}$ to evaluate the overall fit quality of each spectrum. For each star, we compare its ${\chi}^2$ value of the fit to a reference value. The latter is obtained by averaging $\chi^2$ of stars located within a small neighborhood in the parameter space spanned by $T_{\rm eff}$, $\log g$, and ${\rm [Fe/H]}$. 
The ${\rm flag}_{{\chi}^2}$ is then defined as
\begin{equation}
    {\rm flag}_{{\chi}^2} \equiv \frac{{\chi}^2}{{\chi}^2_{\rm ref}},
\end{equation}
In this way, stars suffer anomalously poor fits will have large ${\rm flag}_{{\chi}^2}$.

Some spectra exhibit emission features at the centers of strong absorption lines, which can bias parameter estimation and degrade the fit quality. 
To identify such cases, we define a series of line flags ${\rm flag}_{R_{\rm line}}$, expressed as
\begin{equation}
    \mathrm{flag}_{R_{\rm line}}
    =\frac{R_{{\rm line}} - \mathrm{median}(\boldsymbol{R_{\rm line}})}{\mathrm{MAD}(\boldsymbol{R_{\rm line}})}
    \label{equ:flag_Rline}
\end{equation}

\begin{equation}
    R_{\rm line} \equiv \sum_{\lambda \in [\lambda_1,\lambda_2]}
    \Bigl[f_{\rm model}(\lambda;\boldsymbol{\theta}_{\rm adopted}) - f_{\rm obs}(\lambda)\Bigr].
\end{equation}
where $\lambda_1$ and $\lambda_2$ denote the boundary of selected wavelength window. $\{R_{\rm line}\}$ represents the set of residual values computed over the full stellar sample.

This flag is computed for several key spectral features, including H$\alpha$ (6563~\AA), H$\beta$ (4863~\AA), and Ca\,{\sc ii} H\&K (3968~\AA, 3933~\AA). A positive value of ${\rm flag}_{R_{\rm line}}$ greater than 3 may indicate a significant emission line detection. These indices allow users to isolate stars with emission lines for further studies of stellar activity, accretion, or circumstellar environments.

\section{calibration and validation}\label{sec:calibration and validation}

The stellar parameter estimates are subject to systematic errors due to imperfections in the Kurucz spectra models. 
Therefore, external calibration against well-established reference samples is required to place the results on a reliable parameter scale. Such calibration procedures have been commonly adopted in large spectroscopic surveys \citep[e.g.][]{2008AJ....136.2050L, 2008AJ....136.2022L, 2015MNRAS.448..822X}.
Conducting precise calibration across the entire parameter space for our sample is challenging, owing to the lack of golden reference stars.
In this work, we perform an external calibration for FGK stars using high-resolution spectroscopic references, including the large spectroscopic survey datasets of APOGEE DR17 \citep{2022ApJS..259...35A} and GALAH DR3 \citep{2021MNRAS.506..150B}, as well as the PASTEL catalog \citep{2016A&A...591A.118S} and the metal-poor star sample of \citet{2022ApJ...931..147L}. 
Most calibration is performed by linear interpolation of the residual trends. 
Both the raw and calibrated stellar parameters are presented in the released catalog. The calibration mainly removes the systematic trends and does not significantly affect the parameters' random uncertainties.

\begin{figure*}[htp!]
\centering
    \includegraphics[width=0.9\linewidth]{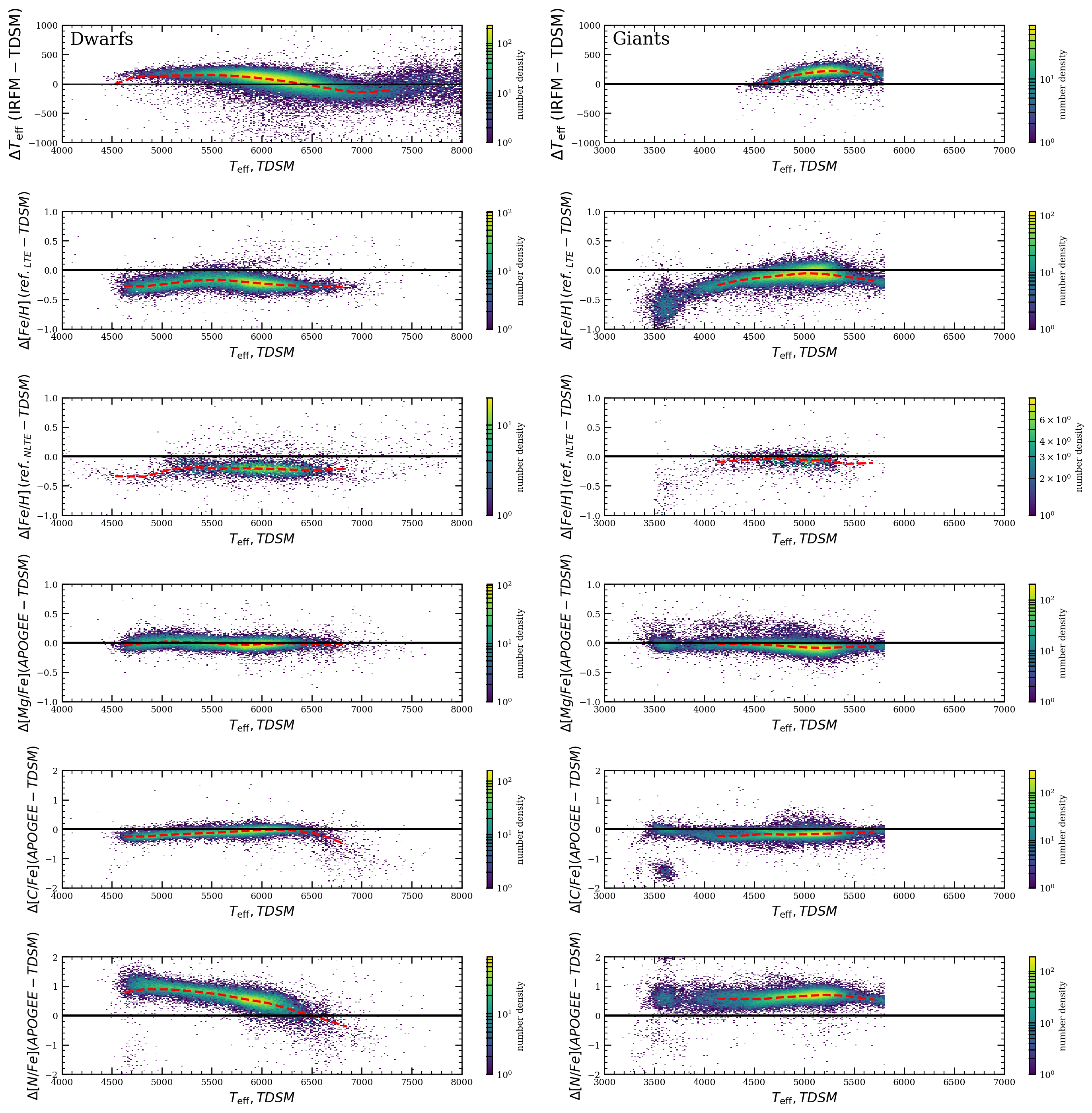}
    \caption{ \label{fig:calibration} Distributions of parameter differences between our estimates and the validation reference results. Results for dwarfs and giants are shown separately. The median trend is shown as a dashed line. }
\end{figure*}

\subsection{Calibration of \texorpdfstring{$T_{\rm eff}$}{Teff}} \label{sec:calibration_teff}

The stellar effective temperature is calibrated to the infrared flux method (IRFM) scale of \cite{2009A&A...497..497G}.
The calibration is performed separately for giant stars and dwarf stars, which are classified based on the raw stellar parameters before calibration. 
Giant stars are defined as $T_{\mathrm{eff}}^{\rm raw}\leq5800$~K and $\log g_{\rm raw}\leq4.5$. 
Here the raw parameters refer to the direct outputs from the spectral model before external calibration, rather than the calibrated final parameters. 
The threshold of $\log g_{\rm raw}\leq4.5$ is adopted because the raw $\log g$ values of cool dwarfs are systematically overestimated in this regime (by more than 0.5 dex). 
Therefore, the commonly adopted solar parameters ($T_{\rm eff}=5777$~K and $\log g=4.44$), which represent the calibrated physical scale in our catalog, are not suitable for this classification. 

As shown in the upper panels of Figure~\ref{fig:calibration}, the $T_{\mathrm{eff}}$ difference exhibits a clear trend with temperature for both dwarfs (upper-left) and giants (upper-right). 
Temperature boundaries are set to 4400–7400~K for dwarfs and 4500–5800~K for giants.
We divide the calibration sample into effective-temperature bins and compute the median difference
\begin{equation}
    \Delta T_{\rm eff}=T_{\rm eff}^{\rm raw}-T_{\rm eff}^{\rm IRFM}
\end{equation}
within each bin. The resulting median trend,\(\Delta T_{\rm eff}(T_{\rm eff})\), is then linearly interpolated as a continuous function of effective temperature. The calibrated temperature is obtained as
\begin{equation}
    T_{\rm eff}^{\rm cal}=T_{\rm eff}^{\rm raw}-\Delta T_{\rm eff}(T_{\rm eff}^{\rm raw})
\end{equation}

For stars outside these ranges, the correction value at the nearest boundary is adopted.

A hard separation between dwarfs and giants would introduce artificial discontinuities at the border. 
To mitigate this effect, we further apply a weighted scheme to stars in the transition region between the dwarf and giant populations, defined by
\(T_{\rm eff}^{\rm raw}<5800\,\mathrm{K}\),
\(\log g_{\rm raw}<4.7\),
and
\(0.0017\,T_{\rm eff}^{\rm raw}-\log g_{\rm raw}>3.8\).
The adopted criterion approximately follows the empirical dwarf--giant separation in the \(T_{\rm eff}\)--\(\log g\) diagram.
An inclined boundary is adopted because systematic offsets in \(\log g\) increase toward cooler dwarfs, making a simple \(\log g\) threshold less effective for selecting giant stars.
Specifically, the offset was calculated as
\begin{equation}
    \Delta T_{\mathrm{eff}} = \frac{(\Delta T_{\mathrm{eff}}^{\rm dwarf} * w_{\rm dwarf} + \Delta T_{\mathrm{eff}}^{\rm giant} * w_{\rm giant})}{w_{\rm dwarf} + w_{\rm giant}},
\end{equation}
where $w_{\rm dwarf}$ and $w_{\rm giant}$ are Gaussian weights centered at $\mu = 5900$~K and $\mu = 5700$~K, respectively, with $\sigma = 100$~K. This ensures a smooth transition between the two calibration schemes.

\subsection{Calibration of \texorpdfstring{$[\mathrm{Fe/H}]$}{[Fe/H]}} \label{sec:calibration_feh}

We provide calibrated metallicity in both LTE and NLTE scales. The two metallicity scales are calibrated independently and are intended for complementary scientific applications. The LTE calibration is tied to APOGEE DR17, PASTEL, and the metal-poor stellar sample of \cite{2022ApJ...931..147L}, while the NLTE calibration is tied to GALAH DR3 and a NLTE version of the metal-poor star sample of \cite{2022ApJ...931..147L}. 
The second and third rows of Figure~\ref{fig:calibration} shows the difference between our [Fe/H] values and the reference as a function of effective temperature. Giants and dwarfs clearly exhibit distinct trends. 

The metallicity calibration is performed in two steps for both dwarfs and giants. Firstly, we compute the median metallicity residual
\begin{equation}
    \Delta_{\rm T}=[\mathrm{Fe/H}]_{\rm raw}-[\mathrm{Fe/H}]_{\rm ref}
\end{equation}
in effective temperature bins and interpolate the resulting trend as a continuous function of \(T_{\rm eff}\). The first-order corrected metallicity is then obtained as
\begin{equation}
    [\mathrm{Fe/H}]_{(1)}=[\mathrm{Fe/H}]_{\rm raw}-\Delta_{\rm T}(T_{\rm eff})
\end{equation}
After removing the temperature dependent systematic trend, the residuals are re-evaluated as a function of metallicity itself. The median residual trend $\Delta_{\rm Fe}([\mathrm{Fe/H}])$ is derived from metallicity bins and interpolated as a continuous function of \([\mathrm{Fe/H}]_{(1)}\). The final calibrated metallicity was obtained as
\begin{equation}
    [\mathrm{Fe/H}]_{\rm cal}=[\mathrm{Fe/H}]_{(1)}-\Delta_{\rm Fe}\!\left([\mathrm{Fe/H}]_{(1)}\right)
\end{equation}

\subsection{Calibration of elemental abundance ratios} \label{sec:calibration_element}
We calibrate the elemental abundance ratios [$\alpha$/Fe], [C/Fe], [N/Fe], and [Mg/Fe] with the values from APOGEE DR17. The calibration derived for [Mg/Fe] is also used to correct for the [$\alpha$/Fe] estimates. The last three rows of Figure~\ref{fig:calibration} shows the differences in [Mg/Fe], [C/Fe], and [N/Fe] for dwarfs and giants. For these abundances, the calibration is performed separately for giants and dwarfs, correcting for systematic trends as a function of effective temperature. Similar to the metallicity calibration, the median residual trend $\Delta_X(T_{\rm eff})$ is derived and interpolated as a continuous function of effective temperature, and the calibrated abundance is obtained by subtracting the corresponding correction from the raw estimate.

\subsection{Calibration and validation of \texorpdfstring{$\log\,g$}{log g}}

\begin{figure*}[t!]
    \centering
    \includegraphics[width=0.95\linewidth]{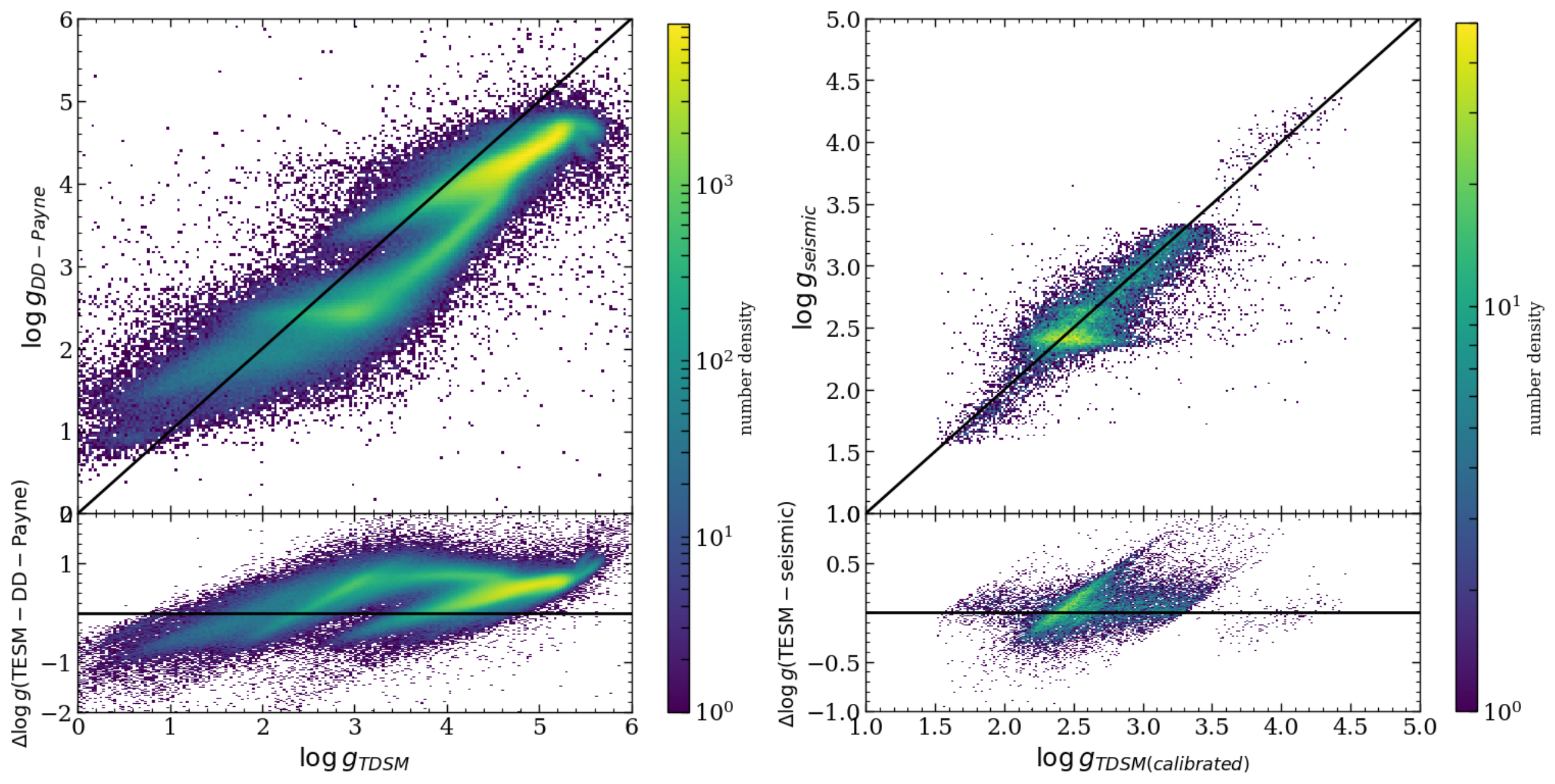}
    \caption{\label{fig:seismology} Validation of $\log~g$ estimates. The left panel shows a comparison of the raw values of $\log~g$ estimates with the \texttt{DD-Payne} results. The right panel shows a comparison of the calibrated $\log~g$ values with asteroseismic estimates in literature  \citep{2018MNRAS.480L..48Y,2022A&A...657A..31M}. The number density of stars is color-coded.}
\end{figure*}

The calibration of surface gravity is more challenging than other parameters. The offsets in $\log~g$ exhibit complex dependencies not only on $\log g$ itself, but also on $T_{\rm eff}$, [Fe/H], and even S/N. A simple interpolation-based calibration, such as that used for $T_{\rm eff}$ and [Fe/H], is insufficient in this case. At the edges of the parameter space or in regimes with sparse cross-matched sources, interpolation becomes unstable, and it may also introduce artificial over-densities along grid boundaries.

To address these issues, we adopted a neural-network–based approach for $\log~g$ calibration. The network takes \{$T_{\rm eff}$, $\log~g$, [Fe/H], S/N\} as inputs and predicts the $\log~g$ offset relative to the reference catalog. This allows the calibration function to flexibly capture the multi-dimensional dependencies of the offset without imposing restrictive functional forms, and ensures smooth behavior across the parameter space. 

We use the DD‑Payne estimates of $\log~g$ for LAMOST DR9 stars \citep{2025ApJS..279....5Z} as the reference values for calibration. This choice is motivated by the broad coverage of the reference sample across the parameter space for both giant and dwarf stars, and by the fact that the DD‑Payne $\log~g$ estimates have been shown to possess high precision and accuracy \citep{2025ApJS..279....5Z}. 

The $\log~g$ calibration is applicable for stars with $3000\ \mathrm{K} \leq T_{\rm eff} \leq 8000\ \mathrm{K}$. For stars cooler than 3000~K or hotter than 8000~K, we adopt the correction value at the respective boundary ($T_{\rm eff}=3000$~K or $T_{\rm eff}=8000$~K) to maintain continuity in the $T_{\rm eff}$–$\log~g$ diagram.

We validate our calibrated $\log~g$ values by comparing them with asteroseismic measurements from the Kepler survey \citep{2010Sci...327..977B}, using the catalogs of \citet{2018MNRAS.480L..48Y} and \citet{2022A&A...657A..31M}. As shown in the right of Figure~\ref{fig:seismology}, the calibrated $\log~g$ values exhibit fairly good consistency with the seismic estimates. 

\subsection{Internal precision from repeated observations}
\begin{figure*}[ht!]
    \centering
    \includegraphics[width=0.9\linewidth]{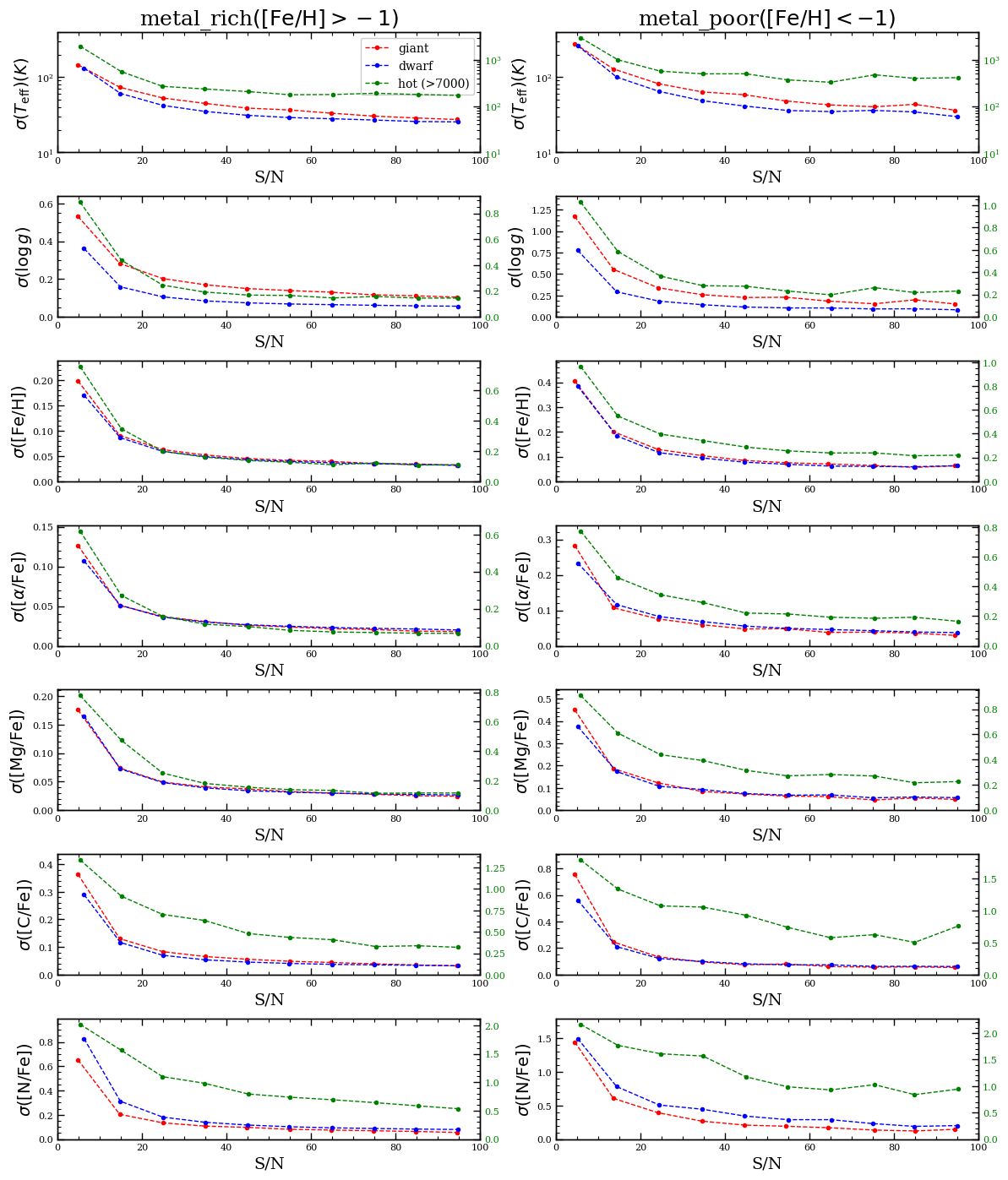}
    \caption{\label{fig:repeat_source} Internal dispersion of stellar parameters deduced from repeated observations, shown for stars with different S/N values. The dispersion was computed using the median absolute deviation (MAD) and converted to a Gaussian-equivalent standard deviation via $\sigma=1.4826 \times MAD$. The left and right columns correspond to results for metal-poor (${\rm [Fe/H]} < -1$) and metal-rich (${\rm [Fe/H]} > -1$) stars, respectively. The red, blue, and green colors represent dwarfs ($T_{\rm eff} < 7500$~K and $\log~g > 4$), giants ($T_{\rm eff} < 6000$~K and $\log~g < 4$), and hot stars ($T_{\rm eff} > 7500$~K), respectively. For the metal-rich sample, the dispersion converges to $\simeq30$~K in $T_{\mathrm{eff}}$, 0.08~dex in $\log g$, 0.04~dex in [Fe/H], 0.02~dex in [$\alpha$/Fe], 0.03~dex in [C/Fe], and 0.05~dex in [N/Fe], at a S/N of $\sim100$. The hot stars have significantly larger dispersion, and their values are shown on the right vertical axis.}
\end{figure*}

We assess the internal precision of our method using repeated observations of the same sources in LAMOST DR10. For each pair of repeat observations, we require the difference in S/N to be less than 10\%, and use the dispersion of the parameter differences as an effective measure of precision. Figure~\ref{fig:repeat_source} shows the dispersion of stellar parameters derived from repeat observations as a function of S/N. Here the dispersion represents the standard deviation after dividing by a factor of $\sqrt{2}$. The results are shown for red giants, FGK dwarfs ($T_{\rm eff}<7500$~K), and OBA stars ($T_{\rm eff}>7500$~K), and for metal-poor (${\rm [Fe/H]}<-1$) and metal-rich (${\rm [Fe/H]}>-1$) stars separately. 

At a S/N=20, the dispersion is $\simeq$51~K in $T_{\rm eff}$, $\simeq$0.13 in $\log~g$, $\simeq$0.07 in $[\mathrm{Fe}/\mathrm{H}]$, $\simeq$0.04 in $[\alpha/\mathrm{Fe}]$, $\simeq$0.06 in $[\mathrm{Mg}/\mathrm{Fe}]$, $\simeq$0.09 in $[\mathrm{C}/\mathrm{Fe}]$ and $\simeq$0.25 in $[\mathrm{N}/\mathrm{Fe}]$ for FGK dwarf stars with ${\rm [Fe/H]}>-1$. 
These values decrease to $\simeq$25~K in $T_{\rm eff}$, $\simeq$0.06 in $\log~g$, $\simeq$0.03 in $[\mathrm{Fe}/\mathrm{H}]$, $\simeq$0.02 in $[\alpha/\mathrm{Fe}]$, $\simeq$0.03 in $[\mathrm{Mg}/\mathrm{Fe}]$, $\simeq$0.03 in $[\mathrm{C}/\mathrm{Fe}]$ and $\simeq$0.08 in $[\mathrm{N}/\mathrm{Fe}]$ at S/N=100. 
The red giants share similar dispersion to the FGK dwarfs except for $\log~g$, which exhibit significant larger values (0.24~dex at S/N=20, 0.10~dex at S/N=100). The OBA stars show systematically larger dispersion for the labels at all S/Ns. Typical values are $\simeq$410 (171) in $T_{\rm eff}$, $\simeq$0.34 (0.14) in $\log~g$, $\simeq$0.27 (0.11) in $[\mathrm{Fe}/\mathrm{H}]$, $\simeq$0.22 (0.07) in $[\alpha/\mathrm{Fe}]$, $\simeq$0.36 (0.12) in $[\mathrm{Mg}/\mathrm{Fe}]$, $\simeq$0.81 (0.32) in $[\mathrm{C}/\mathrm{Fe}]$ and $\simeq$1.33 (0.53) in $[\mathrm{N}/\mathrm{Fe}]$ at S/N=20 (100). 
Similarly, the metal-poor stars also exhibit larger dispersion for all labels, whose values are generally 1.4-2 times that of the metal-rich counterparts. 

Except for $\log~g$, these results are generally consistent with previous works for stars in common parameter domain \citep[e.g.][]{2019ApJS..245...34X, 2022A&A...662A..66X, 2025ApJS..279....5Z}. The $\log~g$ exhibit significantly larger dispersion than that of the DD-Payne estimates \citep[see Figure 9 of][]{2019ApJS..245...34X}. The reason for this difference is unclear. We suspect it might be due to larger uncertainties in the model spectra for red giants.

In ideal case that the spectral flux error is contributed solely by photon noise, one would expect that the dispersion is inversely proportional to the spectral S/N \citep{2017ApJ...843...32T}. However, the real spectra error budget is more complex, considering that systematic errors induced in both spectral modelling and data process contribute a significant portion to the total error budget. At the low-S/N ($<30$) regime, the dispersion is approximately inversely proportional to S/N, while at the high-S/N regime, the dispersion decreases slowly with increasing S/N, and reach a plateau above $S/N\simeq100$.

\subsection{Comparison with high-resolution datasets }

\begin{figure*}[htp!]
    \centering
    \includegraphics[width=0.9\linewidth]{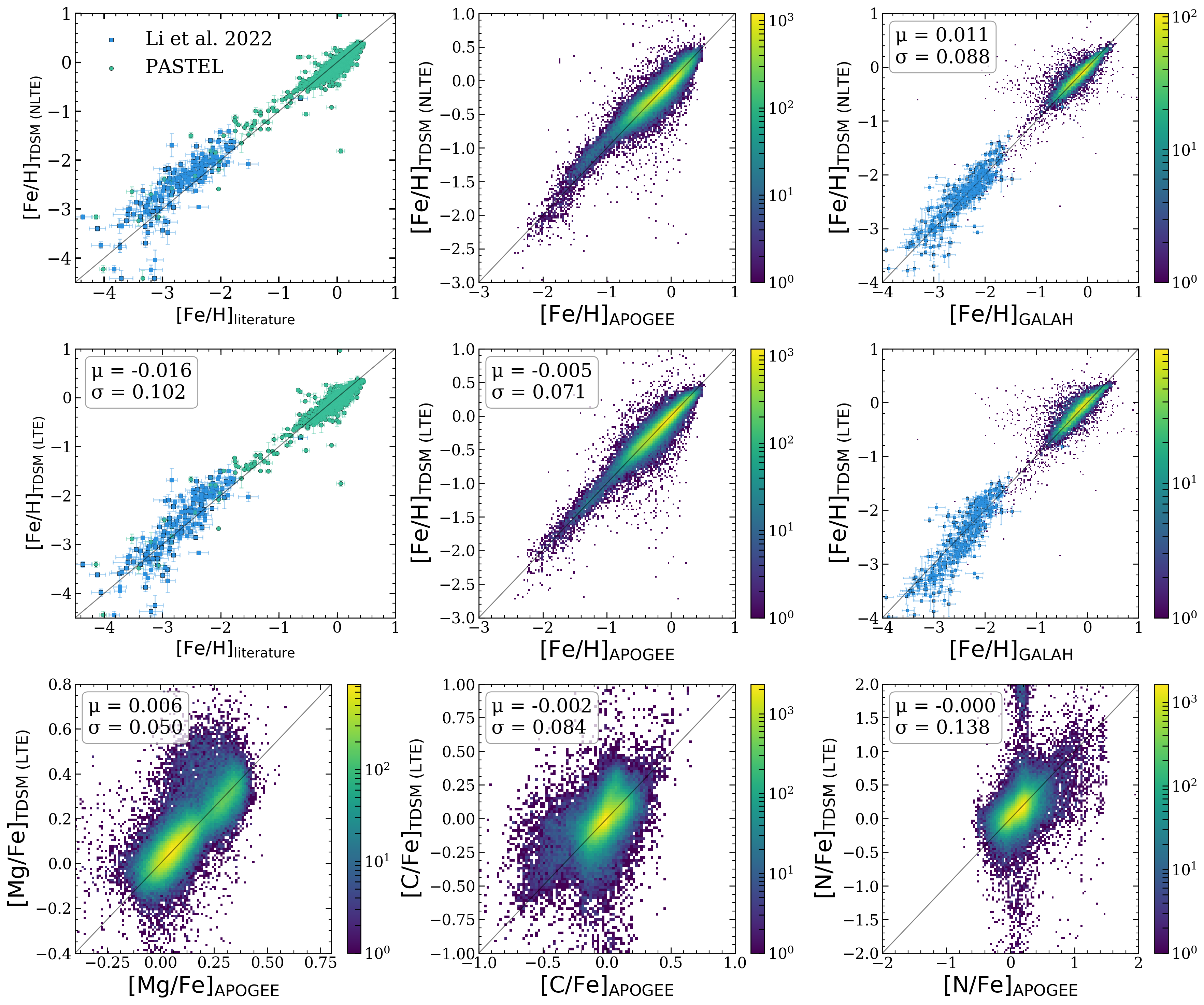}
    \caption{\label{fig:high_resolution} External validation of metallicity and abundance measurements using high-resolution spectroscopic references. The first row shows the comparison of calibrated NLTE metallicity with PASTEL, metal-poor star samples, APOGEE, and GALAH. The second row shows the corresponding comparison using calibrated LTE metallicity. The third row compares [Mg/Fe], [C/Fe] and [N/Fe] with APOGEE. The black line indicates the one-to-one relation. The mean offset and standard deviation are listed in each panel. }
\end{figure*}

Figure~\ref{fig:high_resolution} shows the comparison of the calibrated abundances with the high-resolution reference data sets. Only stars with both $\mathrm{S/N}_g>20$ and $\mathrm{S/N}_i>20$ are shown for the comparison.

The first row compares the calibrated NLTE metallicity with the reference catalogs. The NLTE metallicities show excellent agreement with GALAH DR3, with a scatter of 0.088~dex. However, systematic offsets are evident relative to PASTEL, the metal‑poor sample of \citet{2022ApJ...931..147L}, and APOGEE DR17 at $[\mathrm{Fe/H}]<-1$.

The second row presents the comparison for the LTE metallicity. It agrees  with the combined PASTEL and \citet{2022ApJ...931..147L} sample, as well as with APOGEE DR17, yielding scatters of 0.102~dex and 0.071~dex, respectively. In contrast, larger systematic deviations are observed when compared to references on the NLTE scale.
These comparisons highlight that the LTE and NLTE metallicities provided in our catalog are tied to different reference scales and should therefore be chosen according to the intended scientific application.

The third row compares [Mg/Fe], [C/Fe], and [N/Fe] with APOGEE DR17. The overall consistency is good, with scatters of 0.050, 0.084, and 0.138~dex, respectively. The larger discrepancies in [C/Fe] and especially [N/Fe] near solar abundance ratios are primarily associated with very cool M giants, where molecular features and model‑dependent systematic effects become increasingly important. The additional $T_{\rm eff}$ and $\log g$ comparison with APOGEE could be found in  Appendix~\ref{sec:compare_with_apogee}. 

\section{Stellar parameters of LAMOST DR10}\label{sec:The stellar parameter catalog for LAMOST DR10}

\begin{figure}[htbp]
    \centering
    \includegraphics[width=1 \linewidth]{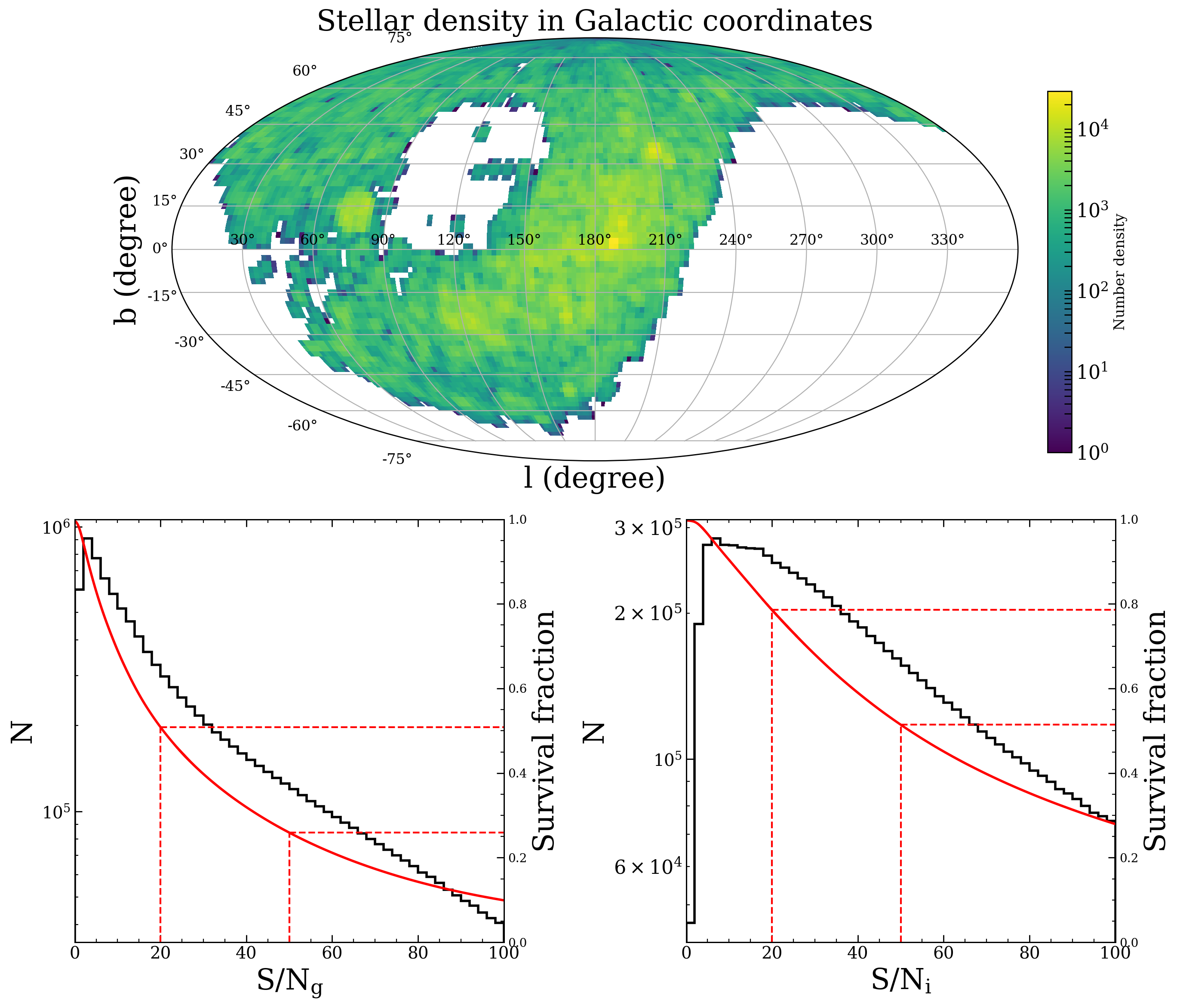}
    \caption{\label{fig:snr+radec} Spatial distribution (top) and spectral S/N distribution (bottom) of the LAMOST DR10 stellar sample. Both the $g-$ and $i$-band spectra S/N are shown. The red dashed lines indicate the cumulative fractions at $\mathrm{S/N}=20$ and $50$.}
\end{figure}

\begin{figure}[htbp]
    \centering
    \includegraphics[width=1 \linewidth]{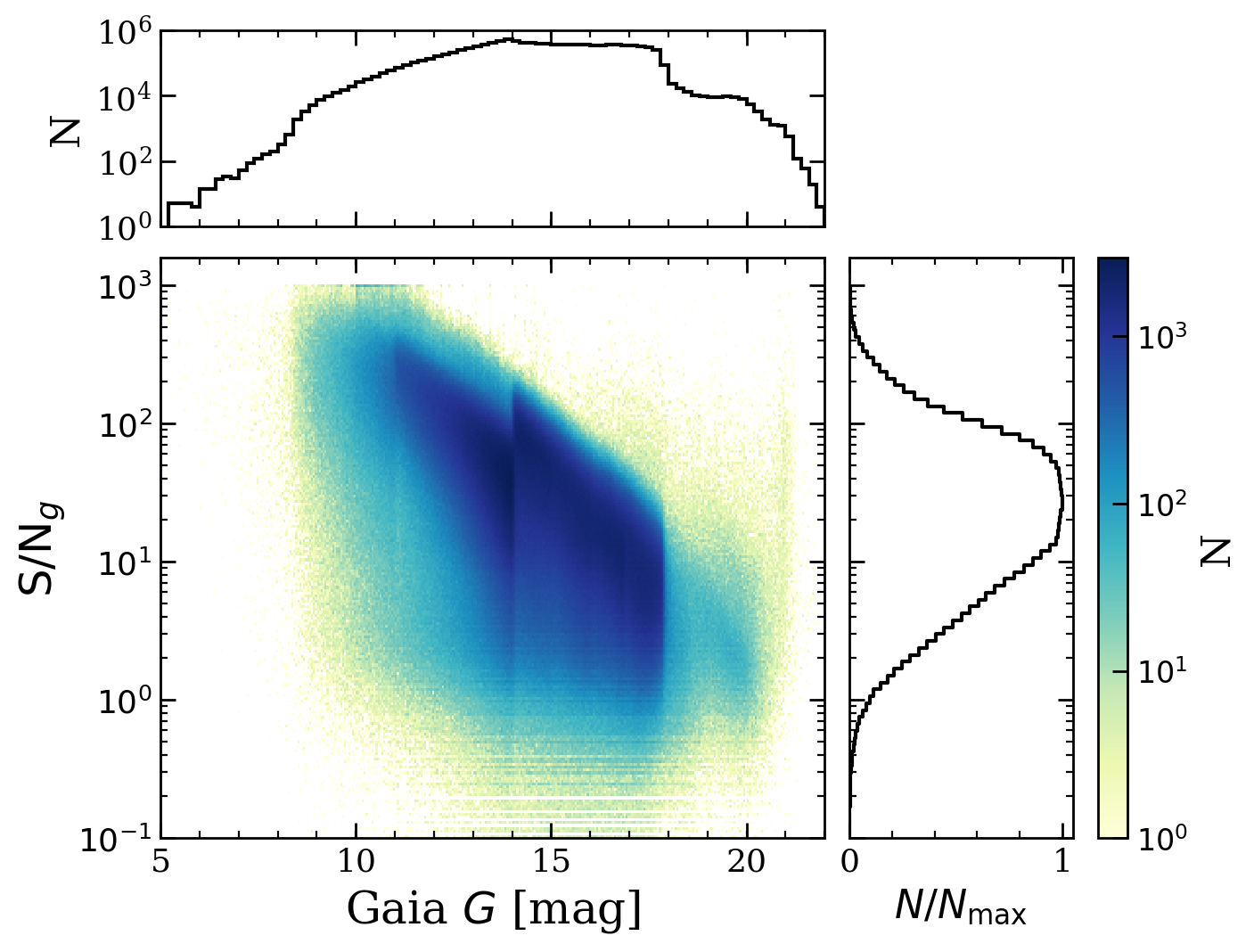}
    \caption{\label{fig:mag_snr} The joint distribution of Gaia \(G\)-band magnitude and \(\log_{10}(\mathrm{S/N}_g)\) for the analyzed LAMOST DR10 spectra. Most targets lie in the range \(G\simeq 9\)--18 mag, and the magnitude distribution peaks at \(G\simeq 14\) mag. The distribution of \(\log_{10}(\mathrm{S/N}_g)\) peaks at around 1.4, corresponding to \(\mathrm{S/N}_g\simeq 25\).}
\end{figure}

The application to LAMOST DR10 results in a catalog containing stellar parameters derived from 11,473,640 low-resolution spectra. The spatial distribution and spectral S/N of the full sample are shown in Figure~\ref{fig:snr+radec}, with S/N computed in the SDSS $g$-band ($\mathrm{S/N}_g$) and $i$-band ($\mathrm{S/N}_i$) wavelength windows. Approximately 51\% (5,835,009 out of 11,473,640) of the spectra have $\mathrm{S/N}_g > 20$, and 26\% (2,974,403) have $\mathrm{S/N}_g > 50$. For the $i$-band, about 79\% (9,015,542) have $\mathrm{S/N}_i > 20$, and 51\% (5,901,905) have $\mathrm{S/N}_i > 50$. Considering both bands simultaneously, roughly 50\% (5,740,860) of the spectra have $\mathrm{S/N} > 20$ in both $g$ and $i$, while 25\% (2,899,594) have $\mathrm{S/N} > 50$ in both bands. 

Figure~\ref{fig:mag_snr} shows the joint distribution of Gaia \(G\)-band magnitude and \(\log_{10}(\mathrm{S/N}_g)\) for the LAMOST DR10 spectra of concern, with marginal histograms shown along the two axes. Most targets lie in the range \(G\simeq 9\)--17.8 mag, and the magnitude distribution peaks at \(G\simeq 14\) mag. 
The distribution of \(\log_{10}(\mathrm{S/N}_g)\) peaks at around 1.4, corresponding to \(\mathrm{S/N}_g\simeq 25\). 
The $g$-band S/N does not represent the spectral quality over the entire wavelength range. Extremely low $\mathrm{S/N}_g$ values are mainly found for cool M dwarfs and M giants, whose flux is weak in the blue region but remains informative in the red wavelength range used for stellar parameter determination.

Table~\ref{table:parameter} summarizes the contents of the catalog. In brief, the catalog includes the sky coordinates of each star, the LAMOST spectrum ID, the stellar parameters derived in this work, and additional auxiliary quantities such as the equivalent widths and line indices described in the preceding sections.

\begin{figure*}[htp!]
    \centering
    \includegraphics[width=0.9\linewidth]{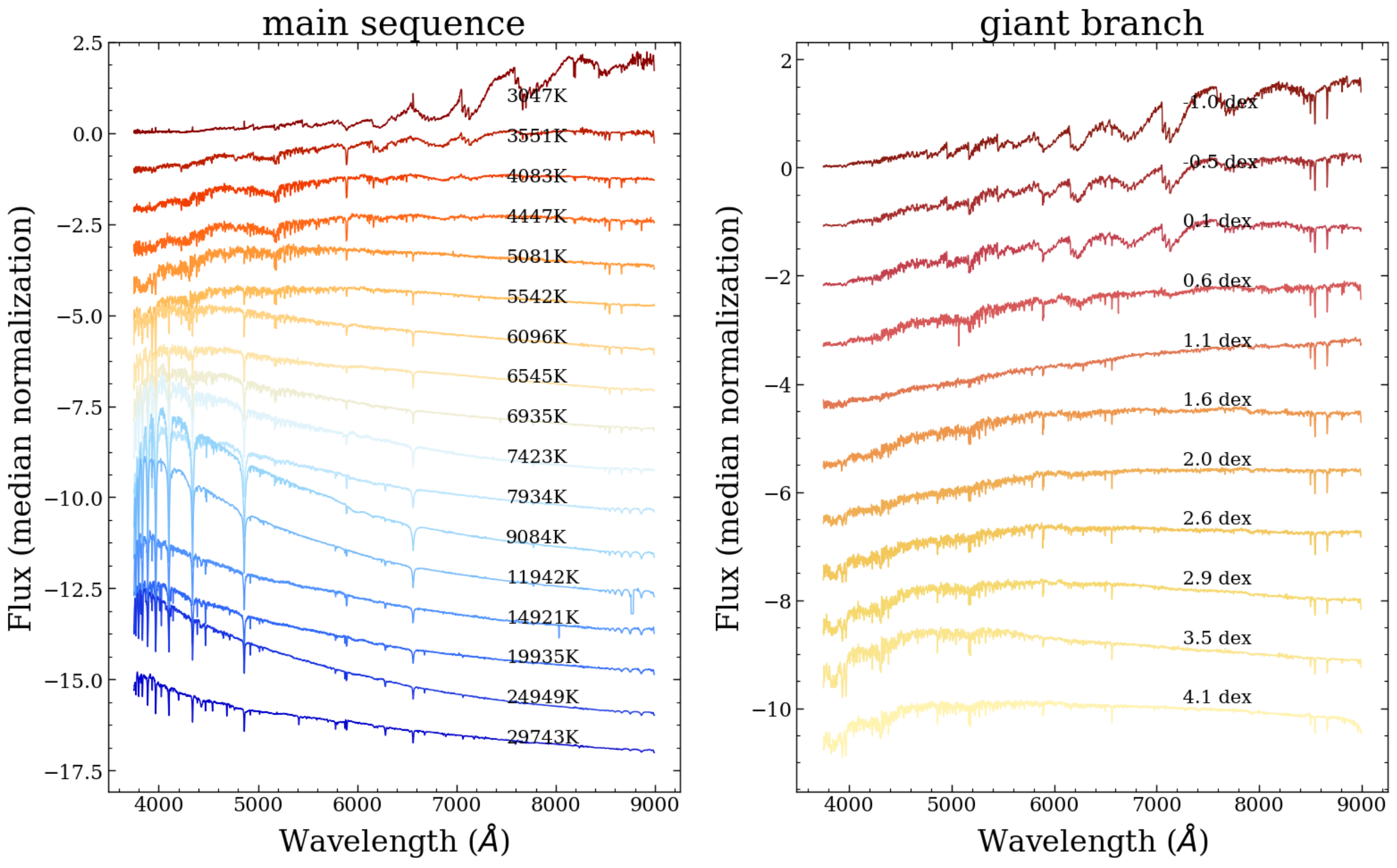}
    \caption{\label{fig:spec_list} 
    Examples of median-normalized LAMOST observed spectra. The left panel shows a sequence of main-sequence stars with different effective temperatures, selected from $\sim$3000\,K to 30000\,K. The right panel shows a sequence of giant-branch stars with different surface gravities, spanning $\log g \sim 4$ to $-1$.}
\end{figure*}

\begin{deluxetable*}{llll}
\tablecaption{Description of the stellar parameter catalog presented in this work.
\label{table:parameter}}
\tabletypesize{\scriptsize}
\tablehead{
\colhead{Column} & \colhead{Format} & \colhead{Unit} & \colhead{Description}
}
\startdata
\texttt{SPECID} & String & -- & ID of this spectrum from LAMOST \\
\texttt{RA} & Float & deg & Right ascension of this source \\
\texttt{DEC} & Float & deg & Declination of this source \\
\texttt{SNR\_G} & Float & -- & Signal-to-noise ratio in g-band\\
\texttt{SNR\_I} & Float & -- & Signal-to-noise ratio in i-band\\
\texttt{FILENAME} & String & -- & Spectrum filename of this source \\
\texttt{TEFF} & Float & K & Effective temperature \\
\texttt{TEFF\_ERR} & Float & K & Error of the effective temperature measurement \\
\texttt{TEFF\_RAW} & Float & K & Uncalibrated effective temperature \\
\texttt{LOGG} & Float & dex & Surface gravity \\
\texttt{LOGG\_ERR} & Float & dex & Error of the surface gravity measurement \\
\texttt{LOGG\_RAW} & Float & dex & Uncalibrated surface gravity \\
\texttt{FEH\_LTE} & Float & dex & LTE scale metallicity \\
\texttt{FEH\_NLTE} & Float & dex & NLTE scale metallicity \\
\texttt{FEH\_ERR} & Float & dex & Error of the [Fe/H] measurement \\
\texttt{FEH\_RAW} & Float & dex & Uncalibrated [Fe/H] \\
\texttt{AFE} & Float & dex & [$\alpha$/Fe] \\
\texttt{AFE\_ERR} & Float & dex & Error of the [$\alpha$/Fe] measurement \\
\texttt{AFE\_RAW} & Float & dex & Uncalibrated [$\alpha$/Fe] \\
\texttt{MG\_FE} & Float & dex & [Mg/Fe] \\
\texttt{MG\_FE\_ERR} & Float & dex & Error of the [Mg/Fe] measurement \\
\texttt{MG\_FE\_RAW} & Float & dex & Uncalibrated [Mg/Fe] \\
\texttt{C\_FE} & Float & dex & [C/Fe] \\
\texttt{C\_FE\_ERR} & Float & dex & Error of the [C/Fe] measurement \\
\texttt{C\_FE\_RAW} & Float & dex & Uncalibrated [C/Fe] \\
\texttt{N\_FE} & Float & dex & [N/Fe] \\
\texttt{N\_FE\_ERR} & Float & dex & Error of the [N/Fe] measurement \\
\texttt{N\_FE\_RAW} & Float & dex & Uncalibrated [N/Fe] \\
\texttt{VMIC} & Float & $\mathrm{km\,s^{-1}}$ & Microturbulent velocity \\
\texttt{VMIC\_ERR} & Float & $\mathrm{km\,s^{-1}}$ & Error of the microturbulent velocity measurement \\
\texttt{CHI2} & Float & -- & Chi-square value of the measurement \\
\texttt{CORR} & Float & -- & Correlation coefficient between the observed spectrum and the model spectrum \\
\texttt{CHI2\_flag} & Float & -- & ${\rm flag}_{\chi^2}$ defined in Section~\ref{data_flag} \\
\texttt{flag\_ew\_Ca\_H} & Float & -- & $\mathrm{flag}_{R_{\rm line}}$ for the Ca\,H line \\
\texttt{flag\_ew\_Ca\_K} & Float & -- & $\mathrm{flag}_{R_{\rm line}}$ for the Ca\,K line \\
\texttt{flag\_ew\_H\_alpha} & Float & -- & $\mathrm{flag}_{R_{\rm line}}$ for the H$\alpha$ line \\
\texttt{flag\_ew\_H\_beta} & Float & -- & $\mathrm{flag}_{R_{\rm line}}$ for the H$\beta$ line \\
\enddata

\end{deluxetable*}

\subsection{The $T_{\mathrm{eff}}$--$\log g$ and $T_{\mathrm{eff}}$--$[\mathrm{Fe}/\mathrm{H}]$ Diagrams}

\begin{figure*}[htp!]
    \centering
    \includegraphics[width=0.9\linewidth]{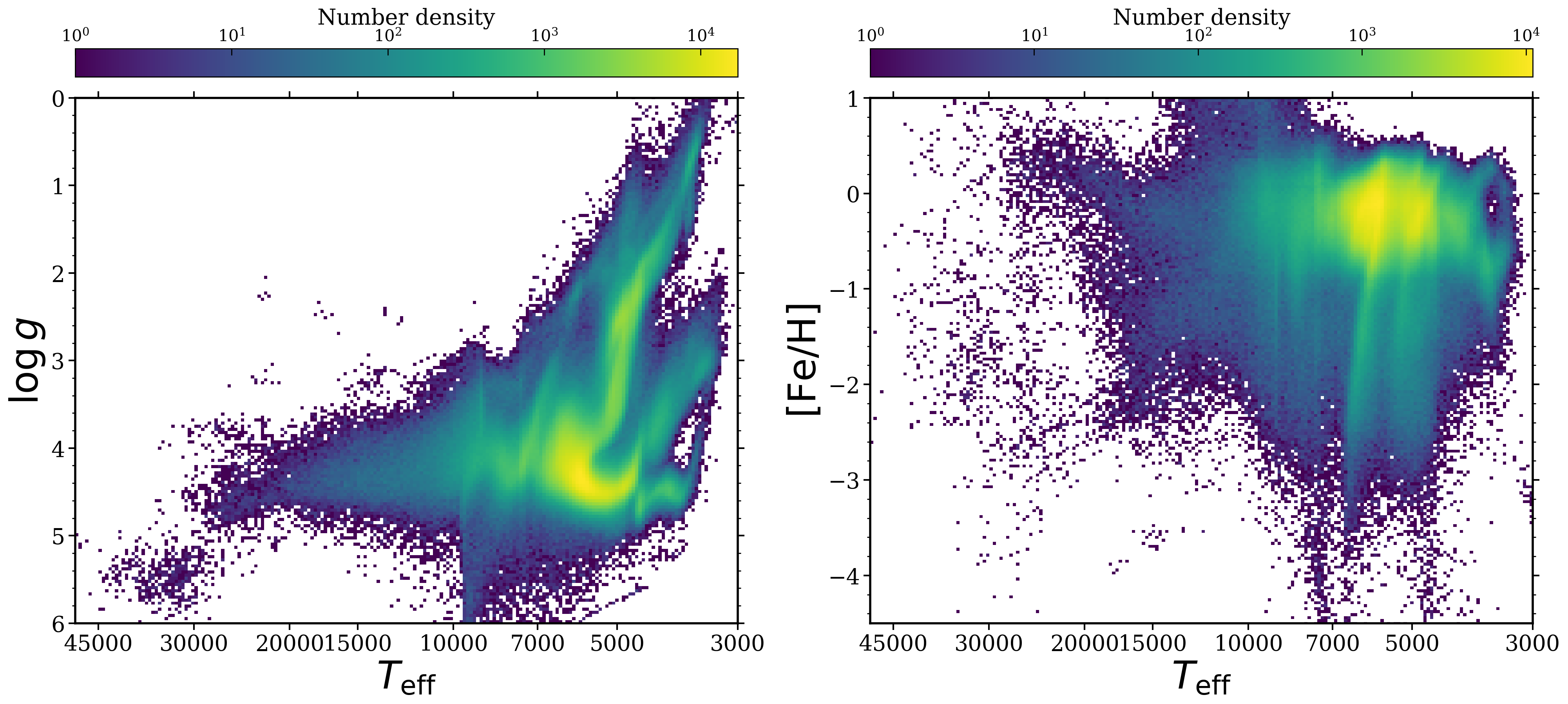}
    \caption{\label{fig:kiel} $T_{\rm eff}$-$\log g$ (left) and $T_{\rm eff}$-$\mathrm{[Fe/H]}$ (right) diagrams of the LAMOST DR10 stellar sample with spectral S/N higher than 20 in both the $g$- and $i$-band, and with ${\rm flag}_{\chi^2}<3$. }
\end{figure*}

The right panel of Figure~\ref{fig:kiel} displays the stellar distribution in the $T_{\rm eff}$-$\log g$ plane (the Kiel diagram) for stars with $\mathrm{S/N}>20$ in both the $g$- and $i$-band wavelength windows, and with ${\rm flag}_{\chi^2}<3$. The sample effectively covers the effective temperature range of $2500<T_{\mathrm{eff}}<45000$~K, the surface gravity range of $-2<\log~g<6$, and the metallicity range of $-4\lesssim{\rm [Fe/H]}\lesssim1$. As an example, Figure~\ref{fig:spec_list} shows some spectra for stars that have different temperature and surface gravity along the main sequence and the giant branch in this broad parameter space.

As shown in the left panel of Figure~\ref{fig:kiel}, FGK dwarfs with $4500 \lesssim T_{\rm eff} \lesssim 7500$~K exhibit the highest stellar density, followed by stars along the red giant branch, particularly the red clump stars at $\log~g \simeq 2.5$. Within the temperature range of FGK stars, metal-poor stars in the red giant and horizontal branch evolutionary phases form distinct sequences. In the $T_{\rm eff}$–[Fe/H] plane, FGK stars display two clear stripes extending from supersolar metallicity down to below $-3.0$ at the metal-poor end. The stripe at higher temperatures exhibits a particularly tight $T_{\rm eff}$–[Fe/H] relation. This stripe corresponds to main-sequence turn-off stars of the Galactic high-$\alpha$ disk, whose temperatures serve as a sensitive indicator of their ages. Consequently, the tight $T_{\rm eff}$–[Fe/H] relation directly reflects the tight age–[Fe/H] relation of the Galactic high-$\alpha$ disk \citep[e.g.][]{2022A&A...662A..66X}.

The hotter OBA stars span from 7500~K to above 30,000~K. Most of them appear as main-sequence dwarfs and subgiants with $3 < \log g < 5$. Above 30,000~K, hot subdwarf stars with $\log g > 5$ dominate the Kiel diagram. In the $T_{\rm eff}$–[Fe/H] plane, OBA stars exhibit a broad range of [Fe/H] values. Metal-poor stars with ${\rm [Fe/H]} < -1$ and $T_{\rm eff} \lesssim 10,000$~K are mainly composed of horizontal branch stars from the Galactic halo and the old (high-$\alpha$) disk. Super-metal-rich stars with ${\rm [Fe/H]} \gtrsim 0.3$ and $7500 < T_{\rm eff} \lesssim 15,000$~K are chemically peculiar stars, predominantly Am/Fm stars \citep[e.g.][]{2016AJ....151...13G,  2020ApJ...898...28X,  2019ApJS..242...13Q,  2023ApJ...943..147S}. For OB stars with $T_{\rm eff} > 15,000$~K, the metallicity distribution shows sparse, clumpy groupings. A prominent feature is a sequence extending from 15,000~K to 30,000~K on the metal-rich side, with roughly solar metallicity but exhibit temperature-dependent trend. These represent the major population of young, main-sequence disk stars. However, a significant number of low-metallicity stars also exist within the same temperature range. The metallicity patterns of these OB stars have not yet been comprehensively characterized.

Note that at around $T_{\rm eff}\simeq10{,}000$~K, artifact vertical stripes appear in both the $T_{\rm eff}$–$\log g$ and $T_{\rm eff}$–[Fe/H] diagrams. These artifacts stem from the imperfect combination of different data sets when accounting for the A-star desert, as described in Section~3.

The cool M-type stars with $T_{\rm eff} \lesssim 4500$~K exhibit a complex distribution in the $T_{\rm eff}$–$\log g$ diagram. The dwarf stars show a clear bifurcation. To our knowledge, such a distinct bifurcation feature in this temperature regime has not been reported in previous studies, and its nature remains unclear. In addition, M-type giant stars also display a bifurcation, with a significant portion showing larger $\log g$ values than the bulk. The origin of this feature is likewise uncertain. The spectra of M-type stars are dominated by molecular bands, which are far less accurately modeled. We therefore caution against the use of the catalog for these stars, particularly for metallicity and abundance measurements.

\subsection{Stellar abundances}
\begin{figure*}[!t]
    \centering
    \includegraphics[width=0.9\linewidth]{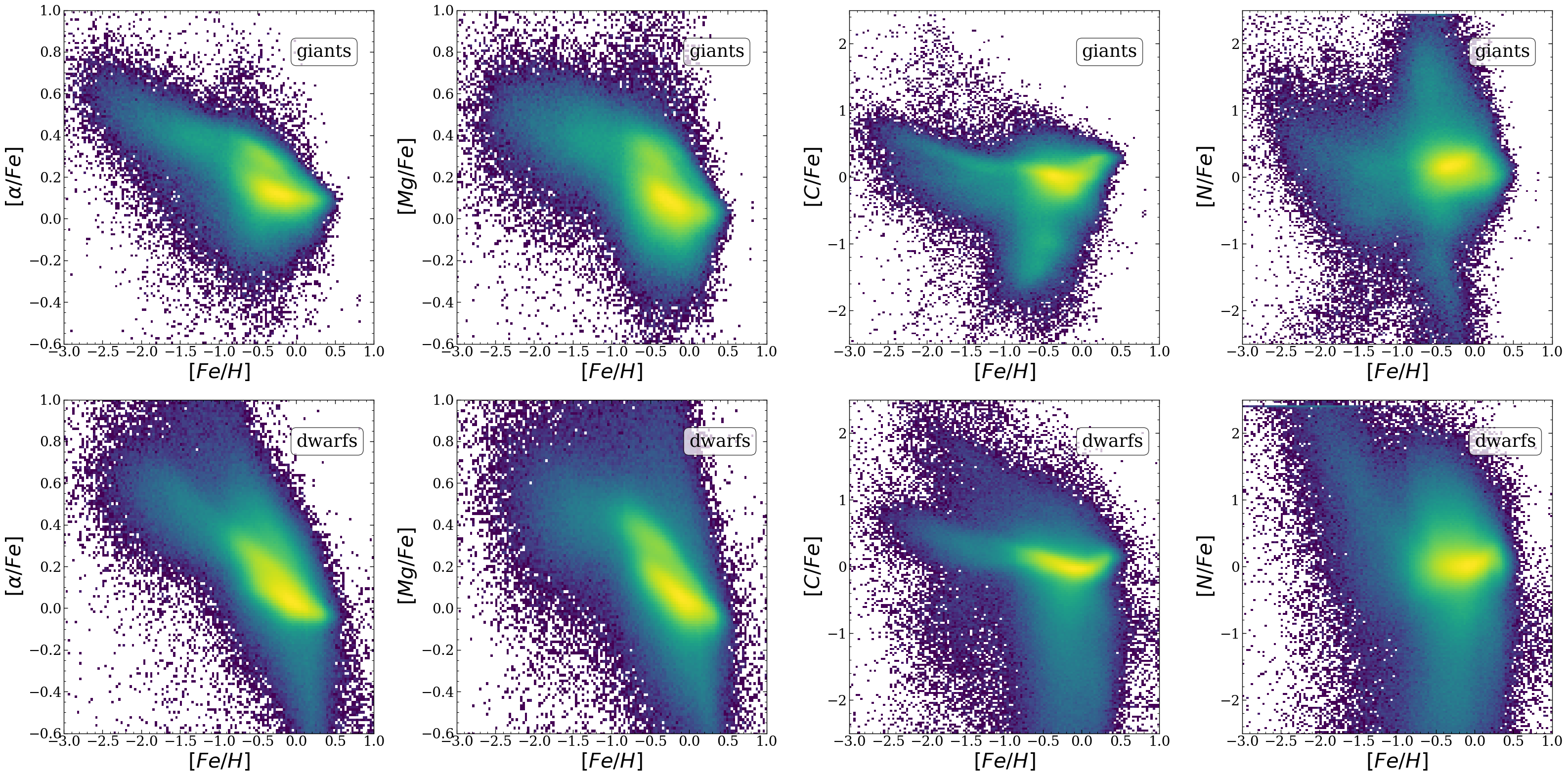}
    \caption{\label{fig:ele_abun} The distribution of element abundances for giant stars ($3800<T_{\rm eff}<5600$\,K, $\log g<3.8$; top panels) and FGK dwarf stars ($4500<T_{\rm eff}<7000$\,K, $\log g>3.8$; bottom panels), selected with either $\mathrm{S/N}_g>50$ or $\mathrm{S/N}_i>50$, and with ${\rm flag}_{\chi^2}<3$.  }
\end{figure*}

The catalog includes the abundance ratios [C/Fe], [N/Fe], [Mg/Fe], and  [$\alpha$/Fe]. Figure~\ref{fig:ele_abun} shows the stellar density distribution in the abundance planes for FGK stars. The bimodal features of the low-$\alpha$ and high-$\alpha$ disks are clearly visible as separate sequences in the [Fe/H]-[$\alpha$/Fe] and [Fe/H]-[Mg/Fe] planes for $\feh \gtrsim -1$. For dwarf stars, this metal-rich regime also hosts a population with very low [Mg/Fe] abundances (${\rm [Mg/Fe]} \lesssim -0.2$). This population consists of chemically peculiar stars, which are found to be ubiquitous among intermediate-mass stars \citep[e.g.][]{2020ApJ...898...28X, 2021MNRAS.506..150B, 2023ApJ...946..110Z}.

For dwarf stars, the majority exhibit a tight [Fe/H]–[C/Fe] sequence, with [C/Fe] approximately solar at solar metallicity, increasing to  ${\rm [C/Fe]}\simeq0.5$ at $\feh=-2$. Giant stars show a similar trend but with a more extended tail of lower [C/Fe] values, as a consequence of the stellar dredge-up process, which decreases the surface C abundance and increases the surface N abundance. Beyond this main sequence, C-rich and C-poor stars are also clearly visible, especially at the metal-poor end. 

Finally, we emphasize that beyond FGK stars, abundance estimates remain poorly calibrated and validated for both hot stars ($T_{\rm eff} \gtrsim 7000$~K) and cold dwarfs ($T_{\rm eff} < 4500$~K). These abundances may be subject to substantial systematic errors due to model imperfections and should therefore be used with great caution.

\subsection{metal-poor stars}
\begin{figure*}[htp!]
    \centering
    \includegraphics[width=0.8\linewidth]{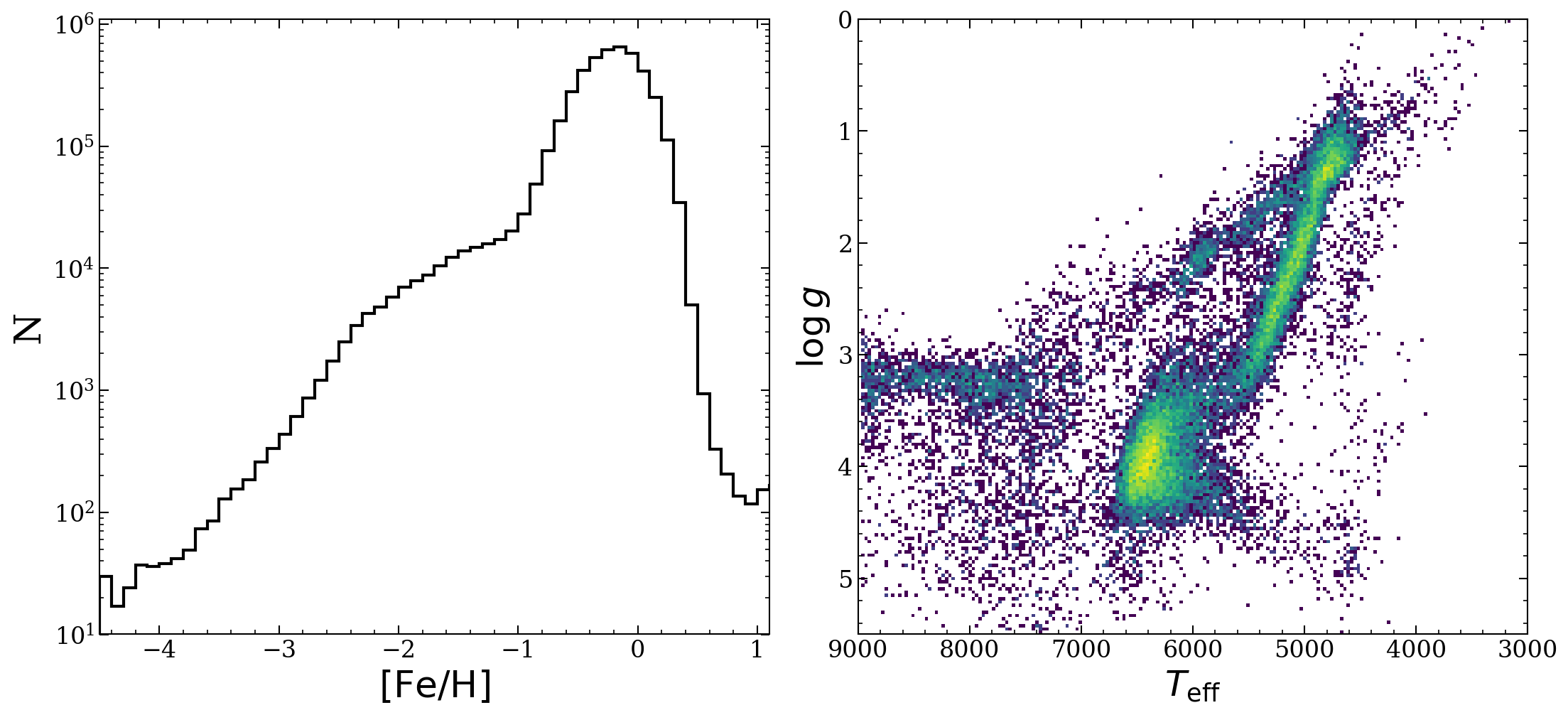}
    \caption{\label{fig:metal_poor} Metallicity distribution of the LAMOST DR10 stellar sample. The left panel shows the overall metallicity distribution for stars with spectral S/N higher than 20 in both the $g-$ and $i-$band wavelength windows, $3000 < T_{\rm eff} < 9000$ K and $-2 < \log g < 5.5$ and ${\rm flag}_{\chi^2}<3$. The right panel shows the $T_{\rm eff}-\log~g$ distribution for very metal-poor stars with $[{\rm Fe/H}] < -2$, highlighting their complete coverage across different evolutionary phases.}
\end{figure*}

The catalog enables a detailed characterization of the metallicity distribution function across a wide parameter space. The left panel of Figure~\ref{fig:metal_poor} shows the metallicity distribution function for all AFGK stars ($T_{\rm eff}<9000$~K) with $S/N_g>20$ and $S/N_i>20$. While the peak number density occurs around ${\rm [Fe/H]}=-0.1$, the distribution exhibits a metal-poor tail extending down to ${\rm [Fe/H]}\lesssim-4$. In total, there are 155,638 metal-poor stars with ${\rm [Fe/H]}<-1$, and 27,321 very metal-poor stars with ${\rm [Fe/H]}<-2$.

As illustrated in the right panel of Figure~\ref{fig:metal_poor}, these very metal-poor stars show a broad distribution in the $T_{\rm eff}$-$\log g$ plane, spanning all evolutionary stages, including the main sequence (MS), main-sequence turn-off (MSTO), subgiant, red giant branch (RGB), and blue horizontal branch. In addition, the figure also clearly reveals the presence of blue straggler stars (BSSs) in the regime with $T_{\rm eff} > 7000\,\mathrm{K}$ and $\log g \gtrsim 4$.

\subsection{Comparison with existing LAMOST catalogs}

\begin{figure*}
    \centering
    \includegraphics[width=0.9\linewidth]{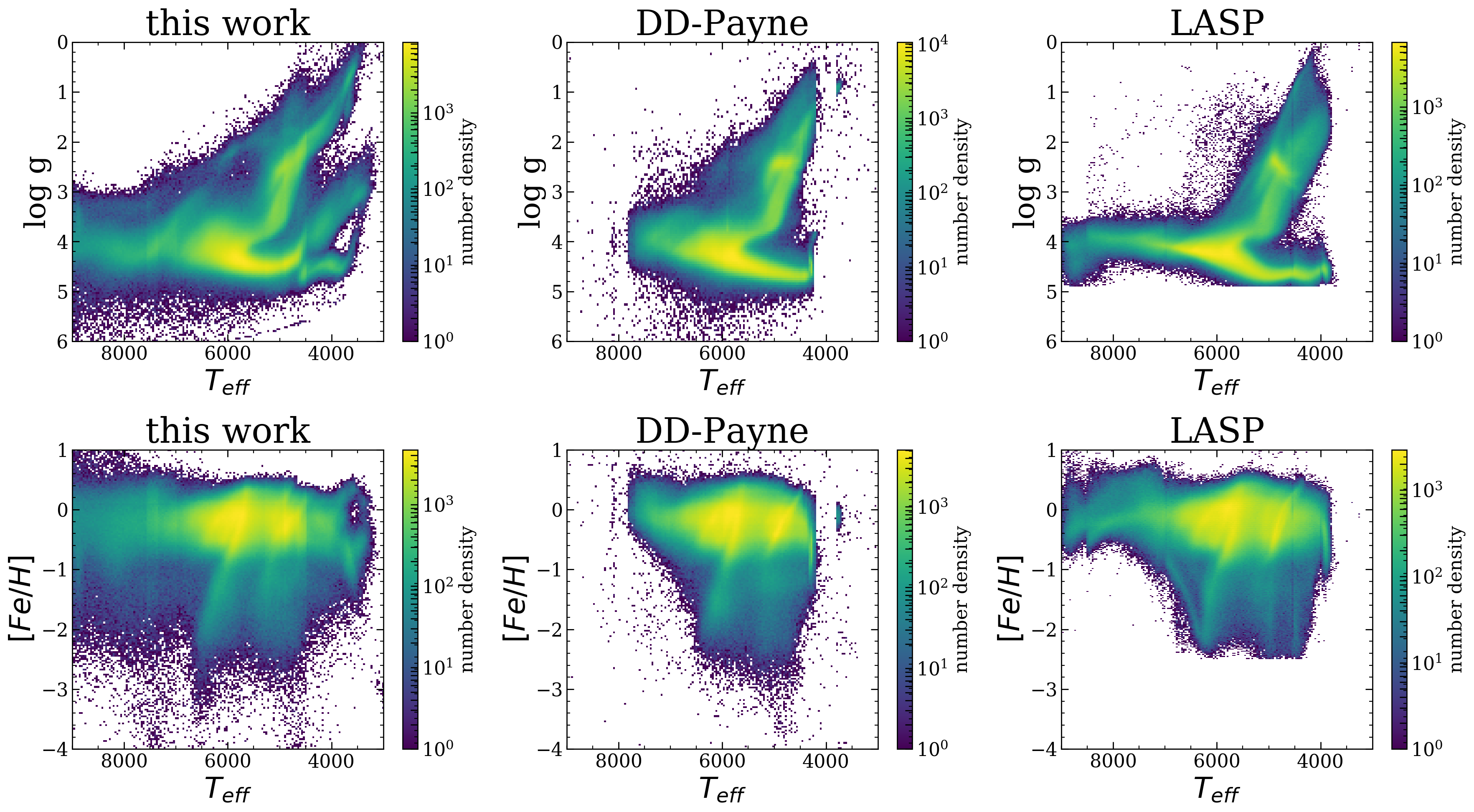}
    \caption{\label{fig:other_catalog1} Comparison of the $T_{\rm eff}-\log~g$ and $T_{\rm eff}-{\rm [Fe/H]}$ diagrams for different LAMOST stellar parameter catalogs. 
    From the left to the right, the results of this work, the DD-Payne estimates \citep{2025ApJS..279....5Z}, and the LASP estimates \citep{2014IAUS..306..340W}  \footnote{http://www.lamost.org/dr10/} are shown, respectively.}
\end{figure*}

We compared the stellar distribution in the $T_{\rm eff}-\log g$ and $T_{\rm eff}-{\rm [Fe/H]}$ planes between our catalog and other LAMOST stellar parameter catalogs, including those derived using LASP \citep{2014IAUS..306..340W} and DD-Payne \citep{2025ApJS..279....5Z}. As shown in Figure~\ref{fig:other_catalog1}, the overall distributions for FGK stars with $4500<T_{\rm eff}<7000$~K, $2<\log g<5$, and ${\rm [Fe/H]}>-2$ show fairly consistent morphologies. Beyond this regime, our catalog exhibits significantly improved parameter coverage over the existing catalogs in several aspects:

\begin{itemize}
\item Our catalog provides rigorous parameter estimates for hotter stars with $T_{\rm eff}>7000$~K, for which the other catalogs are subject to either a hard cut or artificially suppressed distributions, likely due to imperfect templates or algorithms.
\item It reduces biases arising from limitations and selection effects of the training sample. In particular, the model‑driven approach naturally preserves rare stellar populations and unusual objects that might otherwise be suppressed in purely data‑driven analyses.
\item It reveals a distinct horizontal branch in the $T_{\rm eff}-\log g$ plane, while the same sequence appears less clearly defined in other catalogs than in the current work.
\item It uncovers bright red giant stars with $\log g < 1$, which are absent in other catalogs.
\item It extends to metallicities as low as ${\rm [Fe/H]}\simeq-4$, whereas existing catalogs are subject to either a hard cut or a suppressed distribution around ${\rm [Fe/H]}\simeq-2.5$.

\end{itemize}

Given these improvements, our results are expected to be better suited for identifying stars in rare parameter spaces, as well as for statistical studies where artifacts in stellar parameter estimates -- such as suppressed distributions -- could lead to biased results.

\section{Discussion}\label{sec:Discussion}
Model-driven stellar parameter determination is essential to fully exploit the potential of large-scale surveys. Our work has demonstrated the applicability of the TDSM framework for the accurate emulation of low-resolution, blended stellar spectra across an extremely broad parameter space. Within this framework, a lightweight neural network alone is sufficient to achieve high-precision spectral emulation and enables fast, physically sensible determination of stellar parameters for millions of stars. In our case, determining seven stellar labels from 11 million LAMOST spectra ($\sim$4000 pixels per spectrum) required only $\sim$1500 CPU hours on a 2.5~GHz AMD processor. The method can be directly applied to upcoming large surveys, and is particularly valuable for the data analysis of the China Space Station Telescope \citep[CSST, $R\simeq200$;][]{2026SCPMA..6939501C}, Gaia BP/RP spectra \citep{2023A&A...674A...1G, 2023A&A...674A...2D}, and similar surveys with very low spectra resolution.

However, due to imperfections in the ab initio spectral models, the parameter estimates suffer from non‑negligible systematic errors. These errors exhibit complex patterns in the parameter space. We have partly mitigated these systematic errors via external calibration, but future efforts are needed to achieve better calibration, especially for $\log~g$ estimates. One major source of systematic uncertainty in the synthetic spectra arises from the local thermodynamic equilibrium (LTE) assumption adopted in the stellar atmosphere models. Non‑LTE effects are especially severe for warm and hot stars ($T_{\mathrm{eff}} \gtrsim 7000$~K), for which our parameter estimates should be used with caution. Future improvements based on non-LTE atmosphere models such as TLUSTY \citep{1995ApJ...439..875H,2007ApJS..169...83L} may help reduce these systematics. A non‑LTE campaign for hot star parameter estimation is ongoing (Liu et al., in preparation; see also \cite{Aschenbrenner_2026}).

For cool stars ($T_{\rm eff}\lesssim 4000$~K), additional systematic uncertainties may arise from the treatment of molecular opacities, the completeness and accuracy of molecular line lists, and the use of plane-parallel model atmospheres. Future improvements incorporating alternative atmosphere models such as MARCS  \citep{2008A&A...486..951G} and updated molecular line lists may help reduce these systematics in the cool-star regime.

Beyond stellar atmospheric parameters, we have also derived stellar distances, masses, ages, and binary classifications based on the LAMOST spectra and atmospheric parameters presented in this work. These results are presented in a companion paper (Mou et al., in preparation).

\section{Summary}

Stellar parameter determination across a large parameter space is essential for large-scale spectroscopic surveys but remains challenging. In this work, we present \texttt{TDSM}, a physically motivated neural network architecture for self-consistently emulating ab initio stellar spectra -- both the fluxes and their first-order derivatives -- in a high-dimensional parameter space. By applying \texttt{TDSM} to emulate Kurucz model spectra at $R \simeq 2000$, we achieved a typical pixel-wise precision of 0.05\% in flux and $10^{-5}$ to $10^{-4}$ in gradients over the wavelength range $3700<\lambda<9000$~\AA, across the parameter range $3000 < T_{\rm eff} < 50,000$~K, $-1 < \log g < 5$, and $-4.0 < \mathrm{[Fe/H]} < 0.5$.

We have applied \texttt{TDSM} to LAMOST DR10 to derive eight stellar parameters -- $T_{\rm eff}$, $\log g$, $v_{\rm mic}$, [Fe/H], [$\alpha$/Fe], [C/Fe], [N/Fe], and [Mg/Fe] -- from 11 million low-resolution spectra of 8 million unique stars spanning the full model parameter range. The application yields stellar parameter estimates with high internal precision: for spectra with signal-to-noise ratio (S/N) at 50, the typical precision are about 47~K in $T_{\rm eff}$, 0.11~dex in $\log g$, 0.06~dex in [Fe/H], 0.04~dex in [$\alpha$/Fe] and [Mg/Fe], 0.05~dex in [C/Fe], and 0.16~dex in [N/Fe]. 

The parameter estimates are subject to substantial systematic errors, due to imperfections in the underlying ab initio model spectra. To ensure the accuracy of the results, we have performed external calibration to the parameter estimates. This calibration has significantly mitigated systematic errors for FGK stars, but systematic errors for hotter or cooler stars may remain, and their results should therefore be used with caution.

Existing LAMOST stellar parameter catalogs are mainly derived using data-driven approaches, which are limited by their empirical templates or training sets. Compared with these catalogs, our results exhibit significantly better coverage of parameter space, making them especially valuable for identifying stars in rare regions of parameter space, as well as for statistical studies where artifacts in stellar parameter estimates -- such as suppressed distributions -- could lead to biased results. The stellar parameter catalog is made publicly available.

\begin{acknowledgments}
We acknowledge financial support from the National Natural Science Foundation of China (NSFC) grant No. 12588202, the National Key R\&D Program of China grant No. 2022YFF0504200, and NSFC grant Nos. 2022000083 and 12303025.

This work has used spectra data from the Guoshoujing Telescope (LAMOST). LAMOST is a National Major Scientific Project built by the Chinese Academy of Sciences. Funding for the project has been provided by the National Development and Reform Commission. LAMOST is operated and managed by the National Astronomical Observatories, Chinese Academy of Sciences. The LAMOST website is https://www.lamost.org. 

This work has used the ATLAS12 stellar atmospheric model from the Kurucz website (http://kurucz.harvard.edu). We acknowledge Robert L. Kurucz and his team for making their model program publicly available and for their continued efforts in maintaining and updating the model.
\end{acknowledgments}

\appendix

\section{An empirical error model of the LAMOST spectra}\label{sec:residual_model}

We have employed an empirical error model of the LAMOST spectra to perform validation of our spectra model in Section 2.4. Here we introduce how this error model was constructed.

The error model consists of two components: a Gaussian part that accounts for random noise (characterised by the S/N of the spectra) and a non‑Gaussian part that represents systematic errors introduced by instrumental effects, observational limitations, and data‑reduction imperfections. For simplicity, in this work we only consider a constant S/N of 50; for other S/N values, the parameter uncertainties can be scaled using the relation between parameter error and S/N presented in Section 4 (Figure~9).

\begin{figure*}
    \centering
    \includegraphics[width=0.9\linewidth]{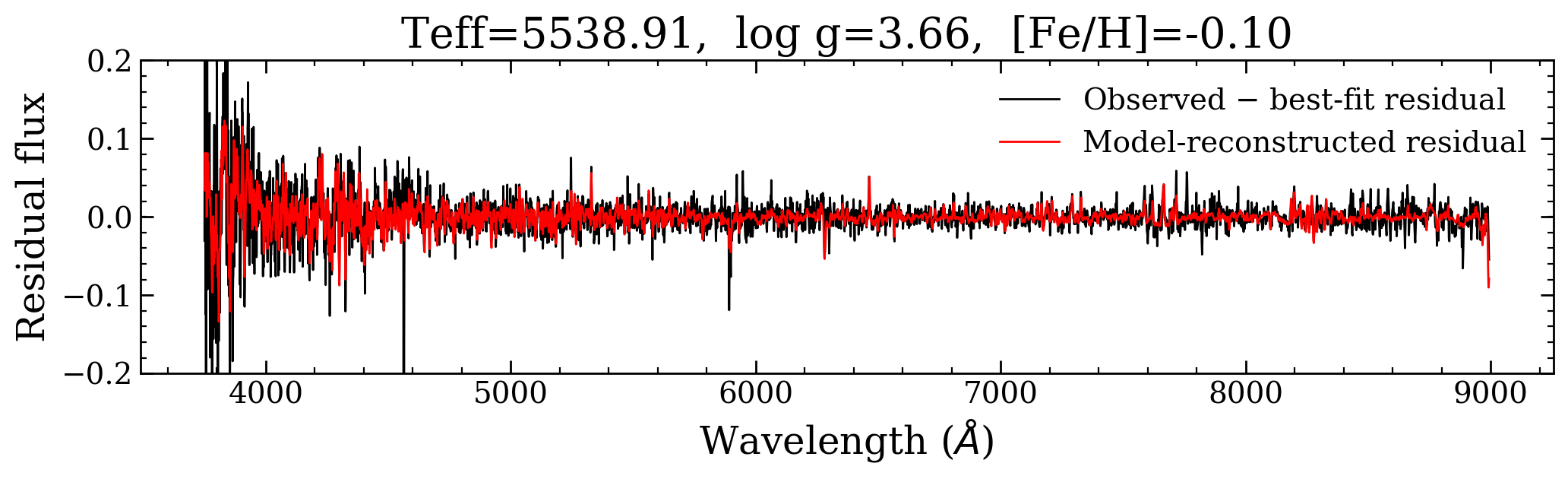}
    \caption{\label{fig:res_model} Comparison between the observed residual spectrum with $SNR_g=49.2$ and the residual reconstructed by the empirical noise model.}
\end{figure*}

The systematic component is modelled from the residuals of spectral fits. Specifically, we randomly selected 40,000 stars satisfying 
$4000<T_{\rm eff}<7000~{\rm K}, 0<\log~g<6, 48<{\rm S/N}_g<52,$ and ${\rm S/N}_i>20$. For each star, we computed the residuals between its LAMOST spectrum and the best‑fitting TDSM model spectrum. We then trained a three‑layer fully connected neural network to capture the parameter‑dependent patterns in these residuals, considering only the fundamental parameters $T_{\rm eff}$,$\log g$, and [Fe/H]. As an example, Figure~\ref{fig:res_model} shows the residual spectrum of one LAMOST star, overlaid with the residuals reconstructed by the neural network model. The reconstructed residuals reproduce the systematic features well while suppressing the random noise. 

It should be noted that the residuals shown in Figure~\ref{fig:res_model} may partly originate from systematic errors inherent in the Kurucz synthetic spectra themselves. Consequently, our approach may overestimate the true systematic errors of the LAMOST spectra, and thus the validation in Section 2.4 may yield conservative (i.e., overestimated) uncertainties in the derived parameters.

\section{Teff and logg comparison with APOGEE} \label{sec:compare_with_apogee}

As described in Section~\ref{sec:calibration and validation}, the $T_{\rm eff}$ and $\log g$ values in this work are calibrated to the IRFM scale and the LAMOST DR9 DD-Payne scale, respectively. Here we present additional comparisons with high-resolution reference datasets. The comparisons are divided into giant stars (top panels) and dwarf stars (bottom panels). Overall, our results show good agreement with the reference measurements. Some systematic deviations are observed for cool giant stars with $T_{\rm eff}<4500$~K, where the calibration becomes less constrained due to the limited availability of reliable reference stars.  In particular, the IRFM temperature scale has limited coverage in this low-temperature regime, which may contribute to the observed deviations. For dwarf stars, the comparison with APOGEE should be interpreted with caution, as the APOGEE temperature scale is also subject to systematic uncertainties in some temperature ranges (e.g., \cite{2025ApJS..279....5Z}). The $\log g$ values show overall good consistency with APOGEE. This agreement is partly expected because our $\log g$ values are calibrated to the DD-Payne scale, whose training labels are largely based on APOGEE measurements. 

\begin{figure*}
    \centering
    \includegraphics[width=0.9\linewidth]{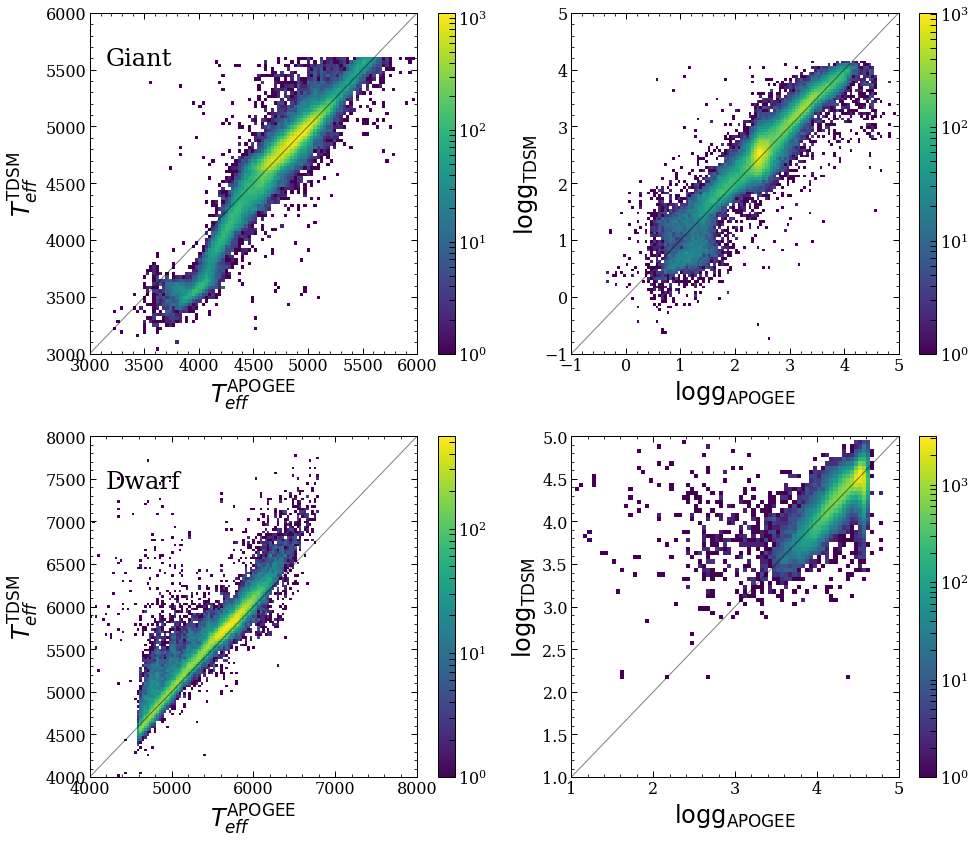}
    \caption{\label{fig:compare_with_apogee}  $T_{\rm eff}$ and $\log g$ results comparison with APOGEE for giant stars (top panels) and dwarf stars (bottom panels).  }
\end{figure*}

\bibliography{sample701}{}
\bibliographystyle{aasjournalv7}

\end{document}